\documentclass[aps,preprint,nofootinbib,amsmath,amssymb,tightenlines,dvips]
               {revtex4}
\usepackage{graphicx}	
\usepackage{bm}		
\usepackage{verbatim}	
\def\prl{{\scriptscriptstyle \parallel}}
\def\prp{{\scriptscriptstyle \perp}}

\def\k{{\bm k}}

\def\x{{\bm x}}
\def\u{{\bm u}}
\def\v{{\bm v}}
\def\J{{\bm J}}
\def\j{{\bm j}}

\def\bpi{{\mbox{\boldmath$\pi$}}}
\def\grad{{\mbox{\boldmath$\nabla$}}}
\def\T{T}
\def\Tr{{\rm Tr}}
\def\vareps{\varepsilon}
\def\vs{v_{\rm s}}
\def\half{{\textstyle {\frac 12}}}
\def\deleps{\delta\vareps}
\def\volume{{\cal V}}
\def\pressure{{\cal P}}
\def\enthalpy{w}
\def\sigmav{\mbox{\boldmath $\sigma$}}
\def\cs{c_{\rm s}}
\def\smallR{{\scriptscriptstyle \mathrm{R}}}
\def\Nc{N_{\rm c}}
\def\coeff#1#2{{\textstyle {\frac {#1}{#2}}}}

\begin{document}

\vspace*{1cm}
\preprint {UW/PT 03-05}

\title {~\kern -20pt
	Hydrodynamic Fluctuations, Long-time Tails, and Supersymmetry
	\kern -20pt\\[20pt]}

\author {Pavel Kovtun\footnotemark\ and Laurence G.~Yaffe\footnotemark}
\def\thefootnote{\fnsymbol{footnote}}
\footnotetext[1]{pkovtun@u.washington.edu}
\footnotetext[2]{yaffe@phys.washington.edu}
\def\thefootnote{\arabic{footnote}}
\affiliation
    {%
    Department of Physics,
    University of Washington,
    Seattle, Washington 98195-1560
    }%

\date{February 28, 2003\\[30pt]}

\begin{abstract}
    Hydrodynamic fluctuations
    at non-zero temperature can cause slow
    relaxation toward equilibrium even in observables which
    are not locally conserved.
    A classic example is the stress-stress correlator in a normal fluid,
    which, at zero wavenumber, behaves at large times as $t^{-3/2}$.
    A novel feature of the effective theory of
    hydrodynamic fluctuations in supersymmetric theories
    is the presence of Grassmann-valued classical fields
    describing macroscopic supercharge density fluctuations.
    We show that
    hydrodynamic fluctuations in supersymmetric theories
    generate essentially the same long-time power-law tails
    in real-time correlation functions that are known in simple fluids.
    In particular, a $t^{-3/2}$ long-time tail must exist in
    the stress-stress correlator of
    $\mathcal{N}=4$ supersymmetric Yang-Mills theory at non-zero temperature,
    regardless of the value of the coupling.
    Consequently, this feature of finite-temperature dynamics
    can provide an interesting test of the AdS/CFT correspondence.
    However, the coefficient of this long-time tail is
    suppressed by a factor of $1/\Nc^2$.
    On the gravitational side, this implies that these long-time tails
    are not present in the classical supergravity limit;
    they must instead be produced by one-loop gravitational fluctuations.
\end{abstract}

\let\tpg=\thepage
\def\thepage{}			
\maketitle

\section{Introduction}
\let\thepage=\tpg		

Holographic AdS/CFT duality implies that properties of
strongly coupled large-$\Nc$ quantum field theories can be
deduced by doing calculations in classical
(super)gravity~\cite{Maldacena,Gubser Klebanov Polyakov,Witten1,AdSCFT Review}.
The correspondence is believed to hold at non-zero temperatures,
and can be helpful in extracting information about
both supersymmetric and non-supersymmetric theories~\cite{Witten2}.
The most well-known example of AdS/CFT duality is $\mathcal{N}{=}4$
supersymmetric Yang-Mills theory with gauge group $SU(\Nc)$,
in $3+1$ dimensions,
which is believed to be dual to type $IIB$ string theory on the
$AdS_5 \times S^5$ background~\cite{AdSCFT Review}.
However, to date, there are very few physical properties which
can be independently calculated on both sides of the duality and
thus used as non-trivial tests of the finite temperature version
of the correspondence.

In this paper, we focus on the low-frequency real-time dynamics
(rather than static thermodynamics) of a finite temperature field theory.
The relevant degrees of freedom are hydrodynamic fluctuations,
by which one means those degrees of freedom whose relaxation time
diverges with wavelength.%
\footnote
    {%
    In normal fluids, the relevant hydrodynamic degrees of freedom
    are fluctuations in energy and momentum density,
    or equivalently temperature and local fluid velocity.
    Hydrodynamic variables can also include fluctuations in
    charge densities associated with any globally conserved currents,
    as well as order parameter phase fluctuations (Goldstone modes)
    in theories with spontaneously broken continuous symmetries.
    }
A suitable version of hydrodynamics is the appropriate form
for an effective theory characterizing these degrees of freedom.
As we will discuss in detail,
one consequence of hydrodynamic fluctuations
at non-zero temperature is the presence of
long-time power-law tails in real-time
correlation functions of conserved
currents~\cite {review of tails,Balescu,Larry and Peter}.
For correlations in the spatial parts of these currents at zero wavenumber,
one finds $t^{-d/2}$ behavior in $d$ spatial dimensions,
or equivalently, non-analytic $\omega^{d/2}$ terms
in the small-frequency behavior of the associated
finite-temperature spectral densities.%
\footnote
    {%
    More precisely, the spectral density divided by $\omega$,
    which is related to the power spectrum of thermal fluctuations,
    has a non-analytic $|\omega|^{(d-2)/2}$ term at low frequency,
    unless $d=4n+2$ in which case an additional $\ln |\omega|$ is present.
    In many respects, long-time hydrodynamic tails are analogous to the
    non-analytic chiral logarithms which appear in one-loop
    chiral perturbation theory (at zero temperature)
    due to quantum fluctuations of massless Goldstone bosons.
    }
These long-time tails are a generic feature of systems which
behave as fluids on arbitrarily long time and large distance scales;
their existence, and the value of the power-law exponent,
are insensitive to the microscopic details of the theory.
Hence, long-time tails in real-time thermal correlators may be
added to the small set of observables about which one
can make firm predictions even in strongly coupled theories.

We will apply our general results to the long-time behavior of thermal
correlation functions in $SU(\Nc)$ $\mathcal{N}{=}4$ supersymmetric
Yang-Mills theory.
Because this is a scale invariant quantum field theory,
it cannot have phase transitions at any non-zero temperature
(in infinite volume).%
\footnote
    {%
    The $\Nc\to\infty$ theory has a phase
    transition in finite volume.
    This is irrelevant for our considerations.
    See Ref.~\cite{Witten2} for more detailed discussion of the phase diagram.
    }
In particular, it must act like a fluid at all non-zero temperatures,
and hence should have a valid hydrodynamic description of its
long-time dynamics.
Consequently, one should (for reasons we will discuss)
expect the zero-wavenumber stress-stress correlator to
show power-law relaxation,
\begin {equation}
    \int d^3\x \>
    \left\langle \left\{ T^{ij}(t,\x), T^{kl}(0) \right\} \right\rangle
    \sim
    t^{-3/2} \,,
\label {eq:stress-tail}
\end {equation}
as $t \to \infty$.

On the gravitational side of the AdS/CFT correspondence,
the stress-stress correlator can be extracted from the
absorption cross-section $\sigma(\omega)$ for scattering of gravitons
by non-extremal three-branes \cite{Klebanov1,Klebanov2}.
This calculation was performed by
Policastro and Starinets \cite{PoliStar}.
In the case when the graviton frequency $\omega$ was small,
Policastro, Son and Starinets \cite{Son et al} interpreted
the zero-frequency limit as the shear viscosity
of $\mathcal{N}=4$ supersymmetric Yang-Mills
plasma.
In Ref.~\cite{PoliStar}, the
small frequency behavior of the cross-section was found to have the form
$\sigma(\omega)=\sigma(0)+O(\omega^2)$.
However, if the stress-stress correlator does exhibit the
power-law tail (\ref{eq:stress-tail}), then AdS/CFT duality
implies that the corresponding graviton
cross-section should be non-analytic at small $\omega$
and behave as
$\sigma(\omega) - \sigma(0) \sim |\omega|^{1/2}$.

This apparent contradiction is the motivation for this paper.
Our goal is to determine the coefficient of the long-time tail
in the stress-stress correlator (\ref {eq:stress-tail}), and in
analogous correlators involving spatial parts of other conserved
currents.
To do so, we will have to construct the correct effective description
of low-frequency, long-distance dynamics in supersymmetric theories.
We will find that supersymmetry (or superconformal)
invariance of a quantum field theory does not preclude the
existence of long-time tails, in accord with
hydrodynamic expectations.
However, in large $\Nc$ gauge theories,
we will find that
a $1/\Nc^2$ suppression appears in
the amplitude of long-time tails.
Thus, on the gravity side of the correspondence, the
small-frequency non-analyticity can not be seen in classical
supergravity (considered in Ref.~\cite{PoliStar}), but should emerge
from a one-loop calculation on a plane-symmetric AdS black hole background.

The paper is organized as follows.
Section~\ref{sec:general HD} reviews what is meant by an effective hydrodynamic theory
and summarizes the properties of hydrodynamic fluctuations in
typical high temperature relativistic theories.
Included is a discussion of why hydrodynamics predicts that
fluctuations in the stress tensor $\T^{ij}$, at zero wavenumber,
decay as $t^{-3/2}$ at long times due to their coupling to both sound waves
and transverse momentum density fluctuations.
In Section~\ref{sec:susy HD} we construct effective hydrodynamics for
supersymmetric theories,
and argue that long-time tails are necessarily still present.
The treatment of hydrodynamic fluctuations
in sections~\ref{sec:general HD} and \ref{sec:susy HD}
is applicable to general field theories.
In Section~\ref{sec:application to SYM} we specialize the general results
to $\mathcal{N}{=}4$ supersymmetric Yang-Mills theory
in the context of the AdS/CFT correspondence.

\section{Effective hydrodynamics}
\label{sec:general HD}


Hydrodynamics may be viewed as an effective theory describing
the dynamics of a thermal system on length and time scales
which are large compared to any relevant microscopic scale
\cite{Forster,Landau Lifshitz v9}.
It is a classical field theory, whose fields can be
regarded as expectation values in some non-equilibrium thermal ensemble
of microscopic quantum operators
averaged over spatial volumes large compared to microscopic length scales.
The beauty of a hydrodynamic description is that it applies to any
system which acts like a fluid on sufficiently long distance scales,
regardless of the strength of microscopic interactions.
We will be concerned specifically with near-equilibrium behavior and the
resulting dynamics of small perturbations about some equilibrium state.


\subsection {Hydrodynamic variables}
\label{sec:HD variables}

To construct the appropriate effective hydrodynamics for a
particular physical system, one has to identify the relevant
low-frequency degrees of freedom (hydrodynamic variables)
and find the equations of motion which govern their dynamics.%
\footnote
    {%
    Because the effective hydrodynamic theory describes
    dissipative dynamics,
    it is easier to work directly with the equations of motion
    than with a classical Lagrangian formulation.
    }
As in any effective field theory, the symmetries, and symmetry
realizations, of the underlying microscopic theory will constrain
the form of the effective theory.
All remaining dependence
on the microscopic theory will be isolated in the particular
values of some finite set of adjustable parameters appearing
in the effective theory.
In the case of hydrodynamics, these input parameters include
transport coefficients (viscosity, diffusivity, conductivity, etc.)
and equilibrium thermodynamic functions.

The degrees of freedom in a hydrodynamic description will
include a minimal set of fields, whose thermal expectation values distinguish
the possible equilibrium states of the theory.
Among these are the energy and momentum densities,
which we will denote as
\begin {equation}
    \vareps \equiv T^{00} \,,\qquad
    \pi^i \equiv T^{i\,0} \,,
\end {equation}
as well as the charge densities $n_a \equiv J^0_a$
of any other conserved currents $J^\mu_a$.%
\footnote
    {%
    Until otherwise stated, we assume that such symmetry currents are
    ordinary bosonic vector fields ({\em i.e.}, not supercurrents).
    }
The argument for this is as follows.

The usual grand canonical ensemble, with statistical density operator%
\footnote
    {%
    We use a $({-}{+}{+}{+})$ metric convention.
    $\beta \equiv 1/T$ is the inverse temperature,
    $u^\nu$ is the rest frame 4-velocity (satisfying $u^2 = -1$),
    and $\mu_a$ are chemical potentials.
    We assume that translation invariance is a symmetry of the theory,
    and that this symmetry is not spontaneously broken.
    We work in three (flat, infinite) spatial dimensions throughout
    our analysis,
    but all results generalize trivially to $d > 3$ spatial dimensions.
    }
\begin {equation}
\label{eq:density operator}
    \hat \rho
    =
    Z^{-1} \, e^ { \beta \, ( u_\nu P^\nu + \, \mu_a \, N_a ) }
\end {equation}
built from the conserved charges
\begin {equation}
    P^\nu \equiv \int d^3\x \> T^{\nu 0}(\x) \,,\qquad
    N_a \equiv \int d^3\x \> J^0_a(\x) \,,
\end {equation}
describes a manifold of time-independent equilibrium states
in which variations in the thermodynamic parameters $\beta$, $u^\nu$,
and $\mu_a$ produce space-independent variations in the
energy density, momentum density, and other charge densities.
In other words, infinite wavelength variations in these densities
have infinite relaxation time (precisely because they are densities
of conserved charges).
But in any local, causal theory, this implies that arbitrarily long
wavelength variations in these densities must have
relaxation times which diverge with wavelength.
Consequently, fluctuations in conserved charge densities remain
relevant variables on arbitrarily long time scales, and must be
retained in an effective hydrodynamic theory.%
\footnote
    {%
    For theories with approximate symmetries,
    an approximately conserved charge density will only act like a
    hydrodynamic degree of freedom on time scales which are
    short compared to the mean time between charge non-conserving reactions.
    }

If the theory under consideration is in a phase with
spontaneously broken continuous symmetry, then the
phase (or orientation) of the relevant order parameter
is also needed to uniquely characterize equilibrium states,
and fluctuations in the order parameter orientation
will become another hydrodynamic degree of freedom.%
\footnote
    {%
    Despite the pervasive misuse of the phrase
    ``spontaneously broken gauge symmetry'',
    Higgs phases of gauge theories are not examples of
    spontaneous breaking of any physical symmetry
    \cite {complementarity},
    and do not imply the existence of an enlarged manifold of
    physical equilibrium states,
    or the presence of additional hydrodynamic degrees of freedom.
    }
Well known examples include superfluid helium,
where the phase of the condensate wave function
becomes a hydrodynamic variable and is
responsible for the appearance of second sound \cite {Landau Lifshitz v9}.
Another example is the chiral limit of QCD,
whose hydrodynamic variables include the
$SU(N_f)$ orientation of the chiral condensate,
fluctuations in which describe pions~\cite{Son QCD}.
At a second order phase transition, long wavelength
fluctuations in the magnitude of an order parameter acquire
divergent relaxation times (due to critical slowing down),
and also become relevant hydrodynamic degrees of freedom.
Finally, for gauge theories in a Coulomb phase, long wavelength
magnetic fields have divergent relaxation times and must be retained
in a hydrodynamic description;
the effective theory in this case is termed magneto-hydrodynamics.

In the following discussion we will assume, for simplicity,
that all these complications are absent.
That is, we assume that the theory under consideration is in a phase
of unbroken global symmetry, is not in a Coulomb phase, and is not sitting
precisely at a second order phase transition.
Also for simplicity, we will assume that the theory possesses
charge conjugation symmetries under which any global conserved
charges $N_a$ transform non-trivially.
We further assume that the equal-time (or more generally, imaginary-time)
thermal correlation functions of the theory
exhibit a finite correlation length.
This is basically just a restatement of our assumed
unbroken global symmetry, absence of Coulomb phase gauge fields,
and non-critical behavior.
Finally, we assume that the theory under consideration is
an interacting theory which describes a sensible equilibrating
thermodynamic system ---
no discussion of hydrodynamic behavior is applicable
to a non-interacting theory!
All these assumptions hold
in typical field theories without U(1) gauge fields,
at sufficiently high temperatures.%
\footnote
    {%
    This includes non-Abelian gauge theories with most any matter field content,
    because interacting scalar fields, fermions, and
    non-Abelian gauge fields all develop finite
    spatial correlation lengths at high temperature.
    }


\subsection {Constitutive relations}
\label {sec:constitutive}

The resulting hydrodynamic description for this general class of theories
takes the form of exact local conservation laws for the conserved currents,
\begin {eqnarray}
    \partial_\mu \, \T^{\mu\nu}(x) &=& 0 \,,
\label{eq:stress conservation}
\\
    \partial_\mu \, J^\mu_a(x) &=& 0 \,,
\label{eq:current conservation}
\end {eqnarray}
together with constitutive relations,
valid on sufficiently long time and distance scales,
expressing the fluxes of conserved quantities
({\em i.e.}, spatial parts of conserved currents)
as local functionals of the hydrodynamic variables themselves
(the densities of conserved charges).%
\footnote
    {%
    More generally, one may formulate constitutive relations
    which, in addition to terms involving hydrodynamic variables,
    also contain noise terms representing the influence of
    short-distance degrees of freedom on the quantity of interest.
    Such noise terms convert the hydrodynamic equations of motion
    into Langevin equations, and allow the effective hydrodynamic
    theory to generate the correct equal time correlations of
    long wavelength equilibrium fluctuations.
    However,
    these noise terms will not be relevant for our purposes,
    and will be ignored throughout our discussion.
    }
As in any effective theory,
possible terms appearing in these constitutive relations must be
consistent with the symmetries of the underlying theory,
and may be classified according to the number of spatial derivatives,
and powers of fields representing departures from equilibrium.
This is the analogue of the usual power-counting in an effective field theory.

We will focus on the dynamics of fluctuations away from some
charge conjugation invariant equilibrium state,
for which all chemical potentials vanish.
Therefore, the charge density $n_a(x)$ will only be non-vanishing
due to some perturbation away from the equilibrium state of interest.
Similarly, if we work in the rest frame of the system,
then the momentum density $\bpi(x)$ will also be non-vanishing only
due to a departure from our given equilibrium state.

The only terms, linear in deviations from our equilibrium state,
which can appear in the resulting constitutive equation for the charge fluxes
({\em i.e.}, the spatial parts of the conserved currents $J^\mu_a$)
have the form
\begin {equation}
    \j_a = -D_{ab} \, \grad \, n_b \,,
\label {eq:J_a linear}
\end {equation}
where $D = \|D_{ab}\|$ is, in general, a matrix of diffusion coefficients
characterizing, in a linear response approximation,
the flux induced by a spatially non-uniform charge density.
These diffusion coefficients are input parameters to the effective
hydrodynamics.%
\footnote
    {%
    \label {fn:Kubo jj}%
    In terms of equilibrium correlation functions
    in the underlying microscopic theory,
    diffusion coefficients are given by the Kubo formula
    $
	(D\chi)_{ab} = \frac 16 \, \lim_{\omega\to0}
	\int d^4x \> e^{i\omega x^0}
	\left\langle \half \{ J^i_a(x), J^i_b(0) \} \right\rangle
    $,
    where
    $\langle \cdots \rangle = Z^{-1} \mathrm{tr}[e^{-\beta H} \cdots]$
    is an equilibrium thermal average,
    $ \chi = \|\chi_{ab}\| $
    is the charge susceptibility matrix,
    $
	\chi_{ab} =
	\partial n_a / \partial (\beta\mu_b)
	=
	\left[
	\langle N_a \, N_b \rangle - \langle N_a \rangle \langle N_b \rangle
	\right] / {\volume}
    $,
    and $\volume$ is the spatial volume.
    The Kubo formula makes manifest the Onsager relation stating
    that $D\chi$ is a symmetric matrix.
    }
Terms with three or more spatial derivatives are also allowed by symmetry
but, for sufficiently long-wavelength fluctuations,
will be negligible compared to the above single derivative terms.
It should be emphasized that the constitutive relation (\ref{eq:J_a linear})
is not an operator identity, unlike Eq.~(\ref{eq:current conservation}).
Rather it is a relation between expectation values of operators in
non-equilibrium states, valid only when those operators are averaged
over a volume large compared to relevant microscopic scales.

Of course, terms can also appear in the constitutive equation for $\j_a$
which are non-linear in the departure from the given equilibrium state.
At quadratic order%
\footnote{
    Terms higher than quadratic are irrelevant for our purposes.
    This will be discussed further at the end of section
    \ref {sec:stress tails}.
    }
it is possible to add a term of the right symmetry
which involves no spatial gradients, and is proportional to the product
of charge and momentum density fluctuations,
\begin {equation}
    \j_a = -D_{ab} \, \grad n_b + \kappa \, n_a \, \bpi \,.
\label {eq:Jflux}
\end {equation}
Since this new term involves no spatial derivatives,
the coefficient $\kappa$ is completely determined by thermodynamic
derivatives which stay within the manifold of equilibrium states.
Namely, an equilibrium state with non-vanishing chemical potentials,
when viewed in an arbitrary reference frame,
will have an energy-momentum tensor of the perfect fluid form,
\begin {equation}
    T^{\mu\nu}
    =
    (\vareps+\pressure)\, u^\mu u^\nu + \pressure \, g^{\mu\nu}\,,
\label {eq:perfect-fluid}
\end {equation}
and conserved currents
\begin {equation}
    J_a^\mu = n_a \, u^\mu \,,
\end {equation}
where $n_a$, $\vareps$, and $\pressure$ are the
equilibrium charge densities, energy density, and pressure, respectively,
in the rest-frame of the fluid,
and $u^\mu$ is the rest-frame 4-velocity.%
\footnote
    {%
    The local rest frame of any thermal system, at a particular event $x$,
    may always be defined as the frame in which the momentum density $\bpi(x)$
    vanishes.
    The local flow velocity in any other frame
    is then defined as the velocity needed to boost to the local rest frame.
    This implies that the local 4-velocity $u^\nu(x)$ is, in general,
    the timelike eigenvector of $T^\mu_{\;\;\nu}(x)$.
    }
Hence, for infinitesimal boosts away from the rest frame,
the momentum density and charge flux are
related to the flow velocity via
\begin {equation}
    \bpi = (\vareps+ \pressure) \, \v \,,
    \qquad
    \j_a = n_a \, \v \,,
\label {eq:flow-vel}
\end {equation}
implying that the coefficient $\kappa$ appearing in the constitutive
relation (\ref {eq:Jflux}) equals $(\vareps{+}\pressure)^{-1}$.
For later convenience, we will denote the enthalpy density,
which is the sum of energy density and pressure, as
\begin {equation}
    w \equiv \vareps + \pressure \,,
\end {equation}
so $\kappa = w^{-1}$ and $\bpi = w \, \v$.

In the constitutive relation for the spatial stress $T^{ij}$,
terms without spatial derivatives are also completely
determined by equilibrium thermodynamics.
Let $\bar\beta$, $\bar\vareps$, and $\bar\pressure$ denote the
inverse temperature, energy density, and pressure, respectively,
of the chosen equilibrium state with vanishing charge and momentum densities.
Then the stress tensor of a nearby equilibrium state with energy density
$\bar\vareps + \deleps$, momentum density $\bpi$,
and charge densities $n_a$,
expanded to second order in deviations away from the reference state,
has the form
\begin {equation}
    \left. T^{ij} \right|_{\rm equil.}
    =
    \delta^{ij} \left[
	\bar \pressure + \vs^2 \, \deleps
	+ \half \, \xi \, (\deleps)^2
	+ \half \, \Xi_{ab} \, n_a \, n_b
	- \vs^2 \, \frac {\bpi^2}{\bar\enthalpy}
    \right]
    +
    \frac{\pi^i \, \pi^j }{ \bar\enthalpy} \,,
\end {equation}
where $\deleps \equiv T^{00} - \bar\vareps$ is the fluctuation
in energy density,
$\vs \equiv (\partial \bar\pressure/\partial\bar\vareps)^{1/2}$ is the
speed of sound,
$\xi \equiv \partial^2 \bar\pressure/\partial\bar\vareps^2$,
and
$
    \Xi_{ab} \equiv
    \partial^2 \bar\pressure /\partial n_a \partial n_b
$.
(Energy derivatives are to be evaluated at constant charge density,
and vice-versa.)
No terms linear in $n_a$ can be present due to the assumed
charge conjugation invariance of the reference equilibrium state.
An exercise in thermodynamic derivatives shows that
\begin {eqnarray}
    \vs^2
    &=&
	\frac{\partial \bar\pressure }{ \partial \bar\vareps}
    =
	\frac{\partial \bar\pressure }{ \partial \bar\beta}
	\left(
	    \frac{\partial \bar\vareps }{ \partial \bar\beta}
	\right)^{-1}
    =
	\frac{\bar\enthalpy }{ \bar\beta} \, C^{-1} \,,
\\[3pt]
    \xi &=&
    \left[
	(1 + 2\vs^2) \, C^{-1} - \bar\enthalpy \, \Gamma^{\it HHH}
    \right] / \bar\beta
    \,,
\\[5pt]
    \Xi_{ab} &=&
    \left[
	(\chi^{-1})_{ab} - \bar\enthalpy \, \Gamma^{\it NNH}_{ab}
    \right] / \bar\beta
    \,.
\end {eqnarray}
We have defined $C$ and $\chi$
as the mean square fluctuation in
energy or charge per unit volume,
\begin {eqnarray}
    C &\equiv& \overline {\langle H^2 \rangle}_{\rm conn} / \volume \,,
\label {eq:mean square E}
\\
    \chi_{ab} &\equiv&
    \overline {\langle N_a N_b \rangle}_{\rm conn} / \volume \,,
\label {eq:mean square N}
\end {eqnarray}
and $\Gamma^{\it HHH}$ and $\Gamma^{\it NNH}$ as ``amputated''
third order connected correlators of energy and charge,
\begin {eqnarray}
	\Gamma^{\it HHH}
    &\equiv&
	C^{-3} \;
	\overline {\langle H^3 \rangle}_{\rm conn} / \volume \,,
\\
	\Gamma^{\it NNH}_{ab}
    &\equiv&
	C^{-1} \, (\chi^{-1})_{aa'} \, (\chi^{-1})_{bb'} \;
	\overline {\langle N_{a'} \, N_{b'} H \rangle}_{\rm conn} / \volume \,,
\end {eqnarray}
with $\volume$ the spatial volume,
$\overline {\langle ... \rangle}$ denoting an expectation in the
reference equilibrium state,
and $\chi^{-1}$ the matrix inverse
of $\chi \equiv \| \chi_{ab} \|$.
Note that the heat capacity per unit volume
$
    c_v \equiv \partial \vareps / \partial T = \beta^2 \, C
$,
while the charge susceptibility
$
    \partial n_a / \partial \mu_b
    =
    \beta \, \chi_{ab}
$.

Terms linear in momentum density can also appear in the constitutive
relation for stress,
but only when combined with a spatial gradient so as to yield
a rank-2 tensor.
Two linearly independent structures are possible,
proportional to the shear or divergence of the vector field,
so that
\begin {eqnarray}
    T^{ij}
    &=&
    \delta^{ij} \left[
	\bar \pressure + \vs^2 \, \deleps
	+ \half \, \xi \, (\deleps)^2
	+ \half \, \Xi_{ab} \, n_a \, n_b
	- \vs^2 \, \frac{\bpi^2 }{ \bar\enthalpy}
    \right]
    +
    \frac{\pi^i \, \pi^j }{ \bar\enthalpy}
\nonumber\\ && {}
    - \gamma_\zeta \, \delta^{ij} \, \grad\cdot\bpi
    - \gamma_\eta
	\left(
	    \nabla^i \, \pi^j + \nabla^j \, \pi^i
	    - \coeff 23 \, \delta^{ij} \, \grad \cdot \bpi
	\right) .
\label {eq:Tij-std}
\end {eqnarray}
The coefficients $\gamma_\zeta$ and $\gamma_\eta$
are conventionally written as
\begin {equation}
    \gamma_\zeta \equiv \frac{\zeta}{ \bar\enthalpy} \,, \qquad
    \gamma_\eta \equiv \frac{\eta }{ \bar\enthalpy} \,,
\end {equation}
with $\zeta$ and $\eta$ the bulk and shear viscosities, respectively.%
\footnote
    {%
    \label {fn:Kubo viscosity}%
    Kubo formulas for the bulk and shear viscosities are
    $
	\zeta = \frac{\beta}{2} \lim_{\omega\to 0}
	\int d^4x \> e^{i\omega t}
	\left\langle \half \{ \pressure(x), \pressure(0) \} \right\rangle
    $,
    and
    $
	\eta = \frac{\beta }{ 20} \lim_{\omega\to 0}
	\int d^4x \> e^{i\omega x^0}
	\left\langle \half \{ s^{ij}(x), s^{ij}(0) \} \right\rangle
    $,
    where $\pressure \equiv \frac 13 T^{ii}$ and
    $s^{ij} \equiv T^{ij} - \pressure \, \delta^{ij}$
    are the trace and traceless parts of the stress tensor, respectively.
    }

All other terms, consistent with symmetries, which could be added to
the constitutive relations (\ref {eq:Jflux}) or (\ref {eq:Tij-std})
involve either more gradients,
or more powers of fluctuations away from equilibrium
relative to the terms included in (\ref {eq:Jflux}) or (\ref {eq:Tij-std}),
and so are negligible for sufficiently small, long-wavelength fluctuations.%
\footnote
    {
    It turns out that the non-linear interactions of
    hydrodynamic fluctuations cause the coefficients of higher order
    terms to become scale dependent quantities,
    just like couplings in a typical effective field theory
    \cite{higher coefs}.
    }

The above constitutive relation for the stress tensor simplifies
considerably in scale-invariant theories.
The tracelessness of the energy-momentum tensor,
$
    T^\mu_{\;\;\mu} = 0
$,
is an operator identity in such theories,
valid in both equilibrium and non-equilibrium states
(at vanishing chemical potentials).
Consequently, in any scale invariant theory the equation of state is exactly
$\bar\vareps = 3 \bar \pressure$,
the speed of sound $\vs = 1/\sqrt 3$,
and the thermodynamic curvatures $\xi$ and $\Xi$ vanish identically,%
\footnote
    {%
    Hence, for scale invariant theories
    $\bar w = \frac 43 \, \bar\vareps$,
    $C = 4 \, \bar\vareps/\bar\beta$,
    $\Gamma^{\it HHH} = \frac{5 }{ 16} \, \bar\beta / \bar\vareps^2$, and
    $\Gamma^{\it NNH} = \chi^{-1}/\bar\enthalpy$.
    These relations also follow by simple
    thermodynamics, given a free energy
    $F \sim \volume T^4 $ and susceptibility $\chi_{ab} \sim T^3$.
    }
as does the bulk viscosity $\zeta \,$.
Therefore, the constitutive relation for the stress tensor (\ref {eq:Tij-std}) in a
scale invariant theory takes the simpler form
\begin {eqnarray}
    T^{ij}
    &=&
    \delta^{ij}
    \left( \bar \pressure + {\coeff 13} \, \deleps \right)
    +
    \frac{1 }{ \bar\enthalpy} \,
    \left(
	\pi^i \, \pi^j - {\coeff 13} \, \delta^{ij} \, \bpi^2
    \right)
    - \gamma_\eta
	\left(
	    \nabla^i \, \pi^j + \nabla^j \, \pi^i
	    - {\coeff 23} \, \delta^{ij} \, \grad \cdot \bpi
	\right) .
\label {eq:Tij scale-inv}
\end {eqnarray}


\subsection {Linearized Hydrodynamics}

If one retains only terms linear in fluctuations in the
constitutive relations (\ref {eq:Jflux}) and (\ref {eq:Tij-std}),
then it is trivial to solve the resulting linearized hydrodynamic equations.
For charge density fluctuations, inserting $\j = -D \, \grad n$
into the conservation law $\partial_t \, n = -\grad \cdot \j$
yields the diffusion equation $(\partial_t - D \, \nabla^2) \, n = 0$.
(We have suppressed indices distinguishing multiple charge densities,
but in general $D$ is a matrix of diffusion constants.)
Inserting a spatial Fourier transform,
\begin {equation}
    n(t,\x) \equiv \int \frac{d^3\k }{(2\pi)^3} \> e^{i \k\cdot\x} \,
    n(t,\k) \,,
\end {equation}
immediately gives
\begin {equation}
    n(t,\k) = e^{-\k^2 D \, t} \> n(0,\k) \,,
\label {eq:lin-response rho}
\end {equation}
showing that fluctuations with wavevector $\k$ relax diffusively
at a rate $\k^2 D$ which vanishes quadratically as $\k\to 0$.
Momentum density fluctuations may be decomposed into longitudinal and
transverse parts,
\begin {equation}
    \bpi(t,\x) = \int \frac{d^3\k }{ (2\pi)^3} \> e^{i \k\cdot\x}
    \left[\,
	\hat \k \> \pi_\prl(t,\k) + \bpi_\prp(t,\k)
    \right] ,
\end {equation}
where $\hat\k \cdot \bpi_\prp(t,\k) \equiv 0$.
For transverse momentum fluctuations, the conservation relation
$ \partial_t \, \pi^i = -\nabla_j \, T^{ij} $ plus the constitutive
relation (\ref {eq:Tij-std}) again combine to give a diffusion equation,
$ (\partial_t - \gamma_\eta \nabla^2) \, \bpi_\prp = 0$, so that
\begin {equation}
    \bpi_\prp(t,\k) = e^{-\k^2 \gamma_\eta t} \> \bpi_\prp(0,\k) \,.
\label {eq:lin-response pi}
\end {equation}
Because energy and momentum densities are both conserved quantities,
fluctuations in energy and longitudinal momentum density are coupled.
One finds that
both quantities satisfy the damped oscillator equation
$
    (-\partial_t^2 - \gamma_s \, \k^2 \, \partial_t - v_s^2 \, \k^2 ) \,
    \delta\vareps(t,\k) = 0
$,
where $\gamma_s \equiv \gamma_\zeta + \frac 43 \gamma_\eta$,
and that $|\k| \, \pi_\prl(t,\k) = i \partial_t \, \delta\vareps(t,\k)$.
In the long wavelength limit ($\gamma_s |\k| \ll v_s$), one has
weakly damped sound waves,
\begin {equation}\renewcommand {\arraystretch}{0.9}
    \left(
    \begin {array}{c}
	i \, \pi_\prl(t,\k) \\[3pt] v_s \, \delta\vareps(t,\k)
    \end{array}
    \right)
    =
    e^{-\frac 12 \k^2 \gamma_s \, t}
    \left(
	\begin {array}{rr}
	    \cos(|\k|v_s t) & \sin(|\k|v_s t) \\[3pt]
	    -\sin(|\k|v_s t) & \cos(|\k|v_s t)
	\end {array}
    \right)
    \left(
    \begin {array}{c}
	i \, \pi_\prl(0,\k) \\[3pt] v_s \, \delta\vareps(0,\k)
    \end{array}
    \right) ,
\label {eq:lin-response sound}
\end {equation}
propagating at the sound speed $v_s$, and whose energy
attenuates at a rate of $\k^2 \, \gamma_s$.


\subsection {Real-time correlators}

The non-equilibrium linear response results (\ref {eq:lin-response rho}),
(\ref{eq:lin-response pi}) and (\ref{eq:lin-response sound}),
describing the relaxation of specific perturbations away
from equilibrium,
may be converted into equivalent results for real-time
correlation functions characterizing the spectrum of fluctuations
in the equilibrium thermal ensemble.%
\footnote {This is sometimes referred to as Onsager's postulate.
    }
Consider, for example,
the two-point correlation of charge density fluctuations,%
\footnote
    {%
    The effective hydrodynamic theory, in which fields are classical,
    cannot distinguish between the symmetrized correlator
    $\langle \half \{ n(t), n(0)\} \rangle$ and
    the Wightman correlator $\langle n(t) n(0) \rangle$.
    In frequency space, this difference is smaller than the correlators
    themselves by a factor of $\hbar \omega/T$, which is negligible
    in the low frequency domain where the hydrodynamic description is valid.
    We have chosen to write expressions involving the symmetrized
    correlator just because this correlation function is always real
    (for Hermitian operators).
    }
\begin {equation}
    A_{n_a n_b}(t,\k)
    =
    \int d^3\x \> e^{-i \k\cdot \x} \>
    \left\langle \half \left\{ n_a(t,\x), n_b(0,{\bf 0}) \right\}
    \right\rangle \,.
\end {equation}
For $t > 0$, one may insert the linear response result
(\ref {eq:lin-response rho}) and find%
\footnote
    {%
    For $t < 0$, one may use
    time-reversal (or CPT plus rotation invariance)
    to show that $A_{n n}(-t,\k) = A_{n n}(t,\k)$.
    Consequently, a Fourier transform in time gives
    $
	A_{n n}(\omega,\k)
	=
	2\k^2 D \, [ \omega^2 + (\k^2 D)^2]^{-1} \> \chi \,.
    $
    Combined with the Ward identity
    $
	\partial_\mu
	\left\langle \half \{ J^\mu(x), J^\nu(0) \} \right\rangle = 0
    $,
    one finds associated hydrodynamic forms for the charge-current correlator,
    $
	A_{n J^l}(\omega,\k)
	=
	2\omega \, k^l D \, [\omega^2 + (\k^2 D)^2]^{-1} \, \chi
    $,
    and the current-current correlator,
    $
	A_{J^iJ^l}(\omega,\k)
	=
	2\omega^2 \, \delta^{il} D \, [ \omega^2 + (\k^2 D)^2]^{-1} \, \chi
    $.
    As required, this current-current correlator is consistent with
    the Kubo formula of footnote~\ref {fn:Kubo jj}.
    \label {fn:t<0}
    }
\begin {equation}
    A_{n n}(t,\k)
    \equiv
    \|A_{n_a n_b}(t,\k)\|
    =
    e^{-\k^2 D\, t} \, \chi(\k) \,,
\label {eq:rho-rho}
\end {equation}
where $\chi(\k) \equiv \|\chi_{ab}(\k)\|$ is the variance matrix of
equal-time charge density fluctuations,
\begin {equation}
    \chi_{ab}(\k)
    =
    \int d^3\x \> e^{-i \k \cdot \x} \,
    \left\langle \half \left\{ n_a(\x), n_b({\bf 0}) \right\}
    \right\rangle .
\end {equation}
In the small-$\k$ regime where a hydrodynamic description is valid,
one may replace this variance by its small $\k$ limit, which is
the charge susceptibility matrix $\chi$ introduced previously,
\begin {equation}
    \lim_{\k\to0} \chi_{ab}(\k)
    =
    \left[
    \left\langle N_a \, N_b \right\rangle -
    \left\langle N_a \right\rangle \left\langle N_b \right\rangle
    \right] / \volume
    =
    \chi_{ab} \,.
\end {equation}

Analogous results for the real-time correlations of energy and
momentum densities are
\begin{subequations}
\begin {eqnarray}
    A_{\vareps\vareps}(t,\k)
    &=&
    e^{-\frac 12 \k^2 \gamma_s |t|} \, \cos(|\k| v_s t) \; C \,,
\label {eq:eps-eps}
\\[2pt]
    i \, A_{\pi^i\vareps}(t,\k)
    &=&
    \hat k^i \, v_s \,
    e^{-\frac 12 \k^2 \gamma_s |t|} \, \sin(|\k| v_s t) \; C \,,
\label {eq:pi-eps}
\\[2pt]
    i \, A_{\vareps\pi^j}(t,\k)
    &=&
    (\hat k^j / v_s) \,
    e^{-\frac 12 \k^2 \gamma_s |t|} \, \sin(|\k| v_s t) \,
    \left\langle {\coeff 13} {\bm P}^2 \right\rangle / \volume \,,
\label {eq:eps-pi}
\\
    A_{\pi^i\pi^j}(t,\k)
    &=&
    \left[
	( \delta^{ij} - \hat k^i \hat k^j ) \, e^{-\k^2 \gamma_\eta |t|}
	+
	\hat k^i \hat k^j \,
	e^{-\frac 12 \k^2 \gamma_s |t|} \, \cos(|\k| v_s t)
    \right]
    \left\langle {\coeff 13} {\bm P}^2 \right\rangle / \volume \,.
\label {eq:pi-pi}
\end {eqnarray}
\end{subequations}
where
$
    A_{\pi^i\pi^j}(t,\k)
    \equiv
    \int d^3\x \> e^{-i \k\cdot \x} \>
    \left\langle \half \left\{ \pi^i(t,\x), \pi^j(0,{\bf 0}) \right\}
    \right\rangle
$,
etc,
and $C$ is the mean square fluctuation in energy per unit volume
defined in Eq.~(\ref {eq:mean square E}).
The mean-square fluctuation in total spatial momentum may be
deduced from the equilibrium relation (\ref {eq:flow-vel})
between momentum density and flow velocity, which implies
\begin {equation}
    \frac { \left\langle {P}^i {P}^j \right\rangle }{ \volume }
    =
    \frac {1 }{ \beta} \,
    \frac {\partial \pi^i }{ \partial v^j}
    =
    \bar\enthalpy \, T \, \delta^{ij} \,,
\label {eq:mean square P}
\end {equation}
or
$
    \langle \coeff 13 {\bm P}^2 \rangle / \volume
    =
    \bar\enthalpy \, T
$.
Hence
the mean square fluctuations in momentum and energy
are related by the velocity of sound,
\begin {equation}
    \frac {\langle {\coeff 13} {\bm P}^2 \rangle }{ \volume}
    =
    T \, \bar\enthalpy
    =
    T^2 \, \frac{\partial \bar\pressure }{ \partial T}
    =
    \vs^2 \, T^2 \frac{\partial \bar\vareps }{ \partial T}
    =
    \vs^2 \> C \,.
\end {equation}



\subsection{Long-time tails}
\label{sec:stress tails}

Consider a spatial charge flux integrated over all space,
\begin {equation}
    \J_a(t) \equiv \int d^3\x \; \j_a(t,\x) \,.
\end {equation}
If non-linear terms in the constitutive relation for $\j_a$ are ignored,
this zero wavenumber charge flux has no coupling to hydrodynamic variables
due to the gradient in the linear $-D \nabla n$ part of
its constitutive relation.
Hence, a linear response analysis predicts that the zero wavenumber
flux should relax on a short, microscopic time scale.%
\footnote
    {%
    This may be seen explicitly from the linear response form of the
    current-current correlator $A^{il}(\omega,\k)$ shown in
    footnote \ref {fn:t<0},
    whose $\k\to0$ limit is frequency independent,
    $
	\lim_{\k\to0} A^{il}(\omega,\k) = 2 \delta^{il} D \chi
    $.
    }
This conclusion is wrong, because the charge flux does couple to hydrodynamic
degrees of freedom, even at zero wavenumber, through the non-linear
$n \bpi/\bar\enthalpy$ term in its constitutive relation
(\ref {eq:Jflux}).
Although this term is quadratic in deviations away from the chosen
equilibrium state, because it contains no spatial gradients
it will always dominate over the single gradient $-D \nabla n$
term of linear response for sufficiently long wavelengths.

To evaluate the zero-wavenumber
$\langle J_a^i(t) \, J_b^j(0) \rangle$ correlator
correctly for large times,
one needs only to express these currents in terms of the product of charge
and momentum density fluctuations, and then insert the results
(\ref{eq:rho-rho}) and (\ref{eq:pi-pi}) for the spectrum of these fluctuations:
\begin {eqnarray}
    {\volume}^{-1}
    \left\langle \half \{ J_a^i(t), J_b^j(0) \} \right\rangle
    &=&
    \frac{1 }{ \bar\enthalpy^2}
    \int d^3\x \,
    \left\langle
	\half \left\{
	    n_a(t,\x)\pi^i(t,\x), \,
	    n_b(0,{\bf 0})\pi^j(0,{\bf 0})
	\right\}
    \right\rangle
\nonumber\\
    &=&
    \frac{1 }{ \bar\enthalpy^2}
    \int d^3\x \,
    \left\langle
	\half \{ n_a(t,\x), n_b(0,{\bf 0}) \} \vphantom{\pi^i}
    \right\rangle
    \left\langle
	\half \{ \pi^i(t,\x), \pi^j(0,{\bf 0}) \}
    \right\rangle
\nonumber\\
    &=&
    \frac{1 }{ \bar\enthalpy^2}
    \int \frac{d^3\k }{ (2\pi)^3} \>
    A_{n_a n_b}(t,\k) \,
    A_{\pi^i\pi^j}(t,-\k)
\nonumber\\
    &=&
    \frac{T }{ \bar\enthalpy} \> \delta^{ij}
    \int \frac{d^3\k }{ (2\pi)^3} \>
    \left[
	{\coeff 23} \, e^{-k^2 (D+\gamma_\eta)|t|} \chi
	+
	{\coeff 13} \, e^{-k^2 (D+\frac 12\gamma_s)|t|}
	\cos(|\k| v_s t) \, \chi
    \right]_{ab}
\nonumber\\
    &=&
    \frac{T }{ \bar\enthalpy} \> \frac{\delta^{ij} }{ 12 }
    \left\{ \frac{1 }{ [(D {+} \gamma_\eta) \, \pi |t|]^{3/2}} \> \chi
    \right\}_{ab} +
    \hbox {(exponential decay)} \,.
\label {eq:JJtail}
\end {eqnarray}
Factorizing the first correlation function into a product of
two-point correlators is justified because,
in the small-$\k$ hydrodynamic regime, the distribution of
fluctuations is arbitrarily close to Gaussian.%
\footnote
    {%
    The essential point is that hydrodynamic variables represent
    microscopic fields averaged over spatial volumes which
    are large compared to the ``correlation volume'' $\xi^3$
    (where $\xi$ is the static correlation length).
    Hence, a hydrodynamic fluctuation with wavelength $\lambda \gg \xi$
    may be regarded as averaging over roughly $(\lambda/\xi)^3$
    essentially independent correlation volumes.
    By the central limit theorem, the distribution of such an average
    will approach a Gaussian distribution as the number of independent
    samples becomes large.
    Residual non-Gaussian correlations in fluctuations with wavenumbers
    of order $\k$ will generate
    corrections to the result (\ref {eq:JJtail}) which are
    suppressed by an additional factor of $|\k|^3 \xi^3$,
    leading to sub-dominant $O(t^{-3})$ corrections to the
    final result.
    }
After inserting the explicit forms of
$C_{n_a n_b}(t,\k)$ and $C_{\pi^i\pi^j}(t,\k)$ in the third step,
an angular average over the direction of $\k$ was performed.
In the final step,
the term involving longitudinal momentum fluctuations
does not contribute to the $t^{-3/2}$ long-time tail
because of the oscillating cosine; this term only gives rise to an
exponentially falling
$ O[\exp\{-\frac{1}{ 4}|t| v_s^2/(D{+}\half\gamma_s)\}] $ contribution.
The above treatment is essentially perturbative theory for
small fluctuations in the long-lived degrees of freedom;
upon factorization, the two-point correlators of conserved densities
are taken to evolve according to linear response.

A completely analogous long-time tail must also be present in the
zero wavenumber stress-stress correlator due to the
coupling of the stress tensor to products of energy, momentum,
and charge density fluctuations.
Inserting the constitutive relation (\ref {eq:Tij-std}),
and evaluating all the terms which can contribute to a long-time tail
is straightforward, but slightly tedious.
This calculation is summarized in Appendix \ref {app:non-scale-inv}.
Here we will specialize to the simpler case of a scale invariant theory,
for which
\begin {eqnarray}
    T^{ij}(t) &\equiv& \int d^3\x \> T^{ij}(t,\x)
\nonumber\\ &=&
    \int d^3\x \,
	\left\{
	    \delta^{ij}
	    \left[
		\bar\pressure + {\coeff 13} \, \deleps (t,\x)
	    \right]
	    +
	    \frac{1 }{ \bar\enthalpy} \, H^{ij}_{mn} \,
	    \pi^{m}(t,\x) \, \pi^{n}(t,\x)
	\right\} ,
\end {eqnarray}
where
$
    H^{ij}_{mn} \equiv
      \half \, \delta^i_m \, \delta^j_n
    + \half \, \delta^i_n \, \delta^j_m
    - \frac{1}{3} \, \delta^{ij} \, \delta_{mn}
$
is a projector onto traceless symmetric tensors.
Therefore the zero-wavenumber stress-stress correlator in the hydrodynamic
regime is
\begin {eqnarray}
    \volume^{-1}
    \left\langle \half \{ T^{ij}(t), T^{kl}(0) \} \right\rangle
    &=&
    \int d^3\x \,
    \Biggl\{
    \frac{1 }{ \bar\enthalpy^2} \,
    H^{ij}_{mn} \, H^{kl}_{pq}
    \left\langle
	\half \left\{
	    \pi^m(t,\x) \, \pi^n(t,\x), \,
	    \pi^p(0,{\bf 0}) \, \pi^q(0,{\bf 0})
	\right\}
    \right\rangle
\nonumber\\ && \quad \qquad {}
    +
    {\coeff 19} \>
    \delta^{ij} \, \delta^{kl}
    \left\langle
	\half \left\{
	    \vareps(t,\x), \, \vareps(0,{\bf 0})
	\right\}
    \right\rangle
    \Biggr\} \,.
\label {eq:stress-stress-1}
\end {eqnarray}
The first term may be evaluated using the same logic as above.
Retaining only pieces which contribute to the connected correlator
[{\em i.e.}
$
    \langle \half \{ T^{ij}(t), T^{kl}(0) \} \rangle
    -
    \langle T^{ij}(t) \rangle
    \langle T^{kl}(0) \rangle
$],
one finds
\begin {eqnarray}
    &&
    \frac{1 }{\bar\enthalpy^2} \,
    \int d^3\x \;
    H^{ij}_{mn} \, H^{kl}_{pq} \,
    \left\langle
	\half \left\{
	    \pi^m(t,\x)\pi^n(t,\x), \,
	    \pi^p(0,{\bf 0})\pi^q(0,{\bf 0})
	\right\}
    \right\rangle_{\rm conn}
\nonumber\\ && \kern 1.0in {}
    =
    \frac{2 }{ \bar\enthalpy^2} \,
    H^{ij}_{mn} \, H^{kl}_{pq} \,
    \int d^3\x \>
    \left\langle \half \{ \pi^m(t,\x), \pi^p(0,{\bf 0}) \} \right\rangle
    \left\langle \half \{ \pi^n(t,\x), \pi^q(0,{\bf 0}) \} \right\rangle
\nonumber\\ && \kern 1.0in {}
    =
    \frac{2 }{ \bar\enthalpy^2} \,
    H^{ij}_{mn} \, H^{kl}_{pq} \>
    \int \frac{d^3\k }{ (2\pi)^3} \>
    A_{\pi^m\pi^p}(t,\k) \,
    A_{\pi^n\pi^q}(t,-\k)
\nonumber\\ && \kern 1.0in {}
    =
    \frac{2T^2}{15} \,
    H^{ij}_{kl}
    \left[
	\left(\coeff{3}{2}\right)^{3/2} \! + 7 \,
    \right]
    \frac{1 }{ (8 \pi \gamma_\eta |t|)^{3/2}} \, .
\end {eqnarray}
Once again,
contributions which fall exponentially with time have been dropped in the
last step.
The final term in Eq.~(\ref {eq:stress-stress-1}) gives
a time-independent contribution,
\begin {equation}
    {\coeff 19} \> \delta^{ij} \, \delta^{kl}
    \int d^3\x \>
    \left\langle
	\half \left\{
	    \vareps(t,\x), \, \vareps(0,{\bf 0})
	\right\}
    \right\rangle
    =
    {\coeff 19} \> \delta^{ij} \, \delta^{kl} \,
    A_{\vareps\vareps}(t,\k{=}{\bf 0})
    =
    \coeff 49 \> \bar\vareps \, T \, \delta^{ij} \, \delta^{kl} \,.
\end {equation}
This reflects the direct coupling between
stress and fluctuations in energy density.
At zero wavenumber, these become fluctuations in the total energy
(in the equilibrium grand canonical ensemble), which are strictly
conserved in time.
Hence, one finds that the leading long-time behavior of the connected zero-wavenumber
stress-stress correlator in a scale invariant theory is%
\footnote
    {%
    Once again, this result is for three spatial dimensions.
    In $d > 3$ spatial dimensions, the same analysis
    leads to a $|t|^{-d/2}$ long-time tail.
    In two or fewer spatial dimensions, hydrodynamic fluctuations are
    sufficiently infrared-singular that the $\omega\to 0$ limits
    in the Kubo formulas in footnotes \ref {fn:Kubo jj}
    and \ref {fn:Kubo viscosity} fail to exist.
    Instead, one finds that transport coefficients exhibit non-trivial
    scale dependence on arbitrarily long length scales~\cite {2d hydro}.
    }
\begin {eqnarray}
    \volume^{-1}
    \left\langle
	\half \left\{ T^{ij}(t), T^{kl}(0) \right\}
    \right\rangle_{\rm conn}
    &\sim&
    \frac{4}{9} \,
    \delta^{ij} \, \delta^{kl} \, \bar\vareps \, T
\nonumber\\ &+&
    \frac{2}{15}\left[ \left( \coeff 32 \right)^{3/2} \! +7 \, \right]
    \left(
	  \half \, \delta^{ik} \delta^{jl}
	+ \half \, \delta^{il} \delta^{jk}
	- \coeff 13 \, \delta^{ij} \, \delta^{kl}
    \right)
    \frac{ T^2 \, }{ (8 \pi \gamma_\eta |t|)^{3/2} } \,.
\nonumber\\
\label {eq:stresstail-conformal}
\end {eqnarray}

Adding any higher-order terms to the stress tensor
constitutive relation (\ref {eq:Tij scale-inv})
will generate sub-dominant power-law tails in
(\ref {eq:stresstail-conformal})
proportional to $|t|^{-5/2}$, $|t|^{-7/2}$, etc.
This is because such higher-order terms will contain either
additional spatial gradients
or additional powers of fluctuations away from equilibrium.
In either case, one finds contributions which scale with
higher powers of $\k$ as $\k\to0$, implying additional powers of $1/|t|$.


\subsection{Low frequency behavior}

The long-time tails present in the current-current
correlator (\ref {eq:JJtail}) or the stress-stress correlator
(\ref {eq:stresstail-conformal})
automatically imply that the Fourier transforms of these
correlation functions cannot be analytic at zero frequency.%
\footnote
    {%
    This follows from the standard argument showing
    that if an integrable function $\tilde G(\omega)$ is analytic in a strip
    of width $2\delta$ surrounding the real axis, then its Fourier
    transform $G(t)$ falls off exponentially as $O(e^{-\delta |t|})$,
    as may be seen by shifting the contour of integration.
    }
A little analysis shows that if $G(t) \sim \alpha / |t|^{3/2}$
at large times
then%
\footnote
    {%
    More generally, if $G(t) \sim \alpha |t|^{-d/2}$, then
    the leading non-analyticity in $\tilde G(\omega)$ for small frequency is
    $
	\tilde G(\omega) \sim \alpha
	\left[ i^{(d-1)(d+1)/4} \, \sqrt{2}\, \pi / \Gamma(d/2) \right]
	|\omega|^{(d-2)/2}
    $
    for $d$ odd,
    $
	\tilde G(\omega) \sim \, \alpha
	\left[ \pi (-1)^{d/4}/ \Gamma(d/2) \right]
	|\omega|^{(d-2)/2}
    $
    for $d=4n$, and
    $
	\tilde G(\omega) \sim  \, \alpha
	\left[ (-1)^{(d+2)/4} \, 2/ \Gamma(d/2) \right]
	|\omega|^{(d-2)/2} \, \ln|\omega|
    $
    for $d=4n+2$.
    }
\begin {eqnarray}
    \tilde G(\omega)
    &=&
    \int dt \> e^{i \omega t} \> G(t)
    \sim
    -\sqrt{8\pi} \, \alpha |\omega|^{1/2}
    +
    \mbox {(analytic terms)} \,,
\label {eq:non-analytic}
\end {eqnarray}
as $\omega \to 0$.
Consequently, in three spatial dimensions, the Fourier transforms
of the real-time zero-wavenumber correlators
$\left\langle \half \{ J^i_a(t), J^j_b(0) \} \right\rangle$ and
$\left\langle \half \{ T^{ij}(t), T^{kl}(0) \} \right\rangle$
must have square root branch points at
zero frequency whose coefficients are related
via Eq.~(\ref {eq:non-analytic})
to the amplitudes of their long-time tails.

In thermal equilibrium, the spectral density of any two operators $A$ and $B$,
\begin {equation}
    \rho_{AB}(\omega)
    \equiv
    \int dt \> e^{i\omega t} \,
    \left\langle \left[ A(t), B(0) \right] \right\rangle \,,
\end {equation}
is directly related to the
Fourier transform of the corresponding symmetrized correlator
\begin {equation}
    \rho_{AB}(\omega)
    =
    2\tanh (\beta\omega/2)
    \int dt \> e^{i\omega t} \,
    \left\langle \half \{ A(t), B(0) \} \right\rangle \,.
\end {equation}
Consequently, the current-current or stress-stress spectral densities,
divided by $\beta\omega$, must have the same non-analyticity
at $\omega = 0$ as do the Fourier transformed real-time correlators.
For example, in a scale-invariant theory one has
\begin{eqnarray}
\label{eq:Tij spectral density scale-inv}
  \frac{1}{\omega} \int dx^0 \> d^3\x \;
      e^{i \omega x^0} \
      \langle \left[ T_{ij}(x), T_{kl}(0) \right] \rangle
  &=&
       \left(
  	  \half \, \delta_{ik} \delta_{jl}
	  + \half \, \delta_{il} \delta_{jk}
	  - \coeff 13 \, \delta_{ij} \, \delta_{kl}
       \right)
\nonumber\\ && {} \times
      \left[ s^{(0)} + s^{(1)} \,|\omega|^{1/2} + O(|\omega|^{3/2}) \right] ,
\end{eqnarray}
with
\begin{eqnarray}
 s^{(0)} &=&
       4 \eta  \, , \qquad\qquad
 s^{(1)} =
       - \frac{T}{60\pi}\left[ \left( \coeff 32 \right)^{3/2} \!\!+7 \, \right]
      \left(\frac{\bar\enthalpy}{\eta}\right)^{3/2}  \, .
\end{eqnarray}


\subsection{Large $\Nc$ scaling}
\label{sec:large N scaling of tails}

The stress-stress correlator will, of course,
have short time transient contributions in addition to its
long time power-law tail.
In large $\Nc$ gauge theories (or large $N$ matrix models)
if the 't Hooft coupling $g^2 \Nc$ is held fixed as $\Nc$ becomes large,
then the short time contribution to the (connected) stress-stress
correlator will scale as $\Nc^2$,
just because there are $O(\Nc^2)$ degrees of freedom.
In contrast, from Eq.~(\ref {eq:stresstail-conformal}) one sees that
the coefficient of the $t^{-3/2}$ long time tail
will be $O(1)$ as $\Nc \to \infty$
if the shear viscosity divided by the enthalpy, $\gamma_\eta = \eta/w$,
has a finite, non-zero large $\Nc$ limit.
This ratio is effectively a microscopic relaxation time,
the mean free time for large angle scattering of
elementary excitations in the system.
In weakly coupled high temperature gauge theories,
this ratio does have a finite large $\Nc$ limit \cite {AMY leading-log}.
Presumably this is generally true, at least in theories like
$\mathcal{N}{=}4$ supersymmetric Yang-Mills theory which are strongly
believed not to have any phase transition separating their weak and
strong coupling regimes.
Assuming this is the case, the cross-over time (beyond which the long-time
tail dominates over the short time transients) will become arbitrarily large
as $\Nc \to \infty$.%
\footnote
    {
    Completely analogous nonuniformities between the large distance
    and large $N$ limits are well known in other contexts.
    For example, adjoint representation Wilson loops in SU($\Nc$)
    Yang-Mills theory have a cross-over from area-law to perimeter-law
    behavior on a distance scale which diverges as $\Nc \to \infty$.
    And various $SU(N)$ symmetric two-dimensional models have correlators
    which fall with distance like $|x|^{-1/N}$~\cite {Witten}.
    }
For any fixed time $t$, the $t^{-3/2}$ tail will not be visible
in the leading large $\Nc$ behavior of the stress-stress correlator,
but will instead appear as a sub-leading $1/\Nc^2$ relative correction:
$$
 \langle T_{ij}(t) T_{kl}(0) \rangle \sim
 O(N_c^2) \, e^{-t/\tau} + O(1) \, t^{-3/2} \ .
$$
Exactly the same result is true of the long-time tail in current-current
correlators.
Diffusion constants are $O(1)$ as $\Nc\to\infty$ (at least in
weakly coupled hot gauge theories \cite {AMY leading-log})
and therefore for fixed time $t$, the long time tail in the current correlator
(\ref {eq:JJtail}) also scales as $O(1/\Nc^2)$ relative to the short time
transients (which scale the same as the susceptibility $\chi$).

In both cases, the result that long-time tails become sub-leading
effects at large $\Nc$ may be regarded as a consequence
of the fact that equilibrium velocity fluctuations are
``anomalously'' small when $\Nc \to \infty$.
As noted earlier in section \ref {sec:constitutive},
the flow velocity equals the momentum density divided by the enthalpy.
Therefore, the mean square fluctuations in average flow velocity
are directly related to the mean square fluctuations in total momentum.
Using the previous result (\ref {eq:mean square P}), one has
\begin {equation}
    \langle \bar v^i \bar v^j \rangle
    =
    \frac {\langle P^i P^j \rangle}{\bar\enthalpy^2 \, \volume^2}
    =
    \frac {\delta^{ij}}{\bar\enthalpy}  \, \frac {T}{\volume}
    =
    O(\Nc^{-2})
\end {equation}
as $\Nc\to\infty$ for fixed volume.
More generally, this means that the fluctuations in flow velocity averaged
over a patch of fluid of linear size $L$ have a $1/\Nc$ suppression
(on top of the expected $L^{-3/2}$ behavior).
Consequently, in the spatial current density $\j = -D \, \grad n + n \, \v$,
the non-linear $n \, \v$ contribution is $1/\Nc$ suppressed relative
to the linear $-D \, \grad n$ term.
Exactly the same conclusion holds for the terms
quadratic in flow velocity in the stress tensor (\ref {eq:Tij scale-inv})
as compared to the linear terms involving the gradient of the velocity.


\section{Supersymmetric Hydrodynamics}
\label{sec:susy HD}

We now wish to generalize the preceding
discussion of hydrodynamic fluctuations 
to the case of supersymmetric theories.
For ease of presentation,
in this section we consider theories with $\mathcal{N}{=}1$
supersymmetry in four dimensions.
Such theories possess a set of conserved supercurrents,
which we will denote by
$S^\mu_\alpha(x)$ and $\bar S^\mu_{\dot\alpha}(x)\,$,
which are Lorentz vector-spinors.
The corresponding conserved supercharges satisfy
the supersymmetry algebra,%
\footnote{
    We follow the notations
    of Wess and Bagger \cite{Wess and Bagger} for spinors.
    In particular, undotted early Greek indices $\alpha,\beta,$...
    label $(\frac{1}{2},0)$ two-component Weyl spinors,
    while corresponding dotted indices label $(0,\frac{1}{2})$
    conjugate Weyl spinors.
    These indices are raised or lowered using the antisymmetric tensor
    $
	\epsilon_{21}=\epsilon^{12}=
	\epsilon_{\dot 2 \dot 1}=\epsilon^{\dot 1 \dot 2}=1
    $,
    so that
    $
	\xi^\alpha Q_\alpha = - \xi_\alpha Q^\alpha
    $
    and
    $
	\bar\xi^{\dot\alpha} \bar Q_{\dot\alpha} =
	- \bar\xi_{\dot\alpha} \bar Q^{\dot\alpha}
    $.
    Extended Pauli matrices are
    $\sigma^\mu_{\alpha\dot\beta} \equiv (-1,\sigmav)$
    and $\bar\sigma^{\mu\dot\alpha\beta} \equiv (-1,-\sigmav)$, 
    and the spinor representation generators of the Lorentz group are
    $
	(\sigma^{\mu\nu})_\alpha^{\ \beta} \equiv
	\frac{1}{4} \big( 
      \sigma^\mu_{\alpha\dot\alpha} \bar\sigma^{\nu\dot\alpha\beta} - 
       \sigma^\nu_{\alpha\dot\alpha}\bar\sigma^{\mu\dot\alpha\beta} \big)
    $
    and
    $
	(\bar\sigma^{\mu\nu})^{\dot\alpha}_{\ \dot\beta} \equiv
	\frac{1}{4} \big( 
      \bar\sigma^{\mu\dot\alpha\alpha} \sigma^{\nu}_{\alpha\dot\beta} - 
       \bar\sigma^{\nu\dot\alpha\alpha} \sigma^{\mu}_{\alpha\dot\beta} \big)
    $.
    }
\begin {equation}
    \{Q_\beta, \bar{Q}_{\dot\gamma} \} = 2 \sigma_{\beta\dot\gamma}^\mu \,P_\mu
    \,.
\label {eq:susyalg}
\end {equation}
The theory may also have a chiral
$U(1)$ symmetry ($R$-symmetry), whose charge $R$ does not
commute with the supercharges, 
\begin{equation} \label{eq:R charge algebra}
    [Q_\alpha,R]=Q_\alpha\ , \qquad
    [\bar Q_{\dot\alpha},R]=- \bar Q_{\dot\alpha} \,.
\end{equation}
as well as various additional internal
symmetries whose charges $N_a$ do commute with the supercharges.
For definiteness, we will assume that the theory does have $R$-symmetry,
and that this symmetry is not spontaneously broken.%
\footnote
    {%
    In ${\cal N}{=}1$ supersymmetry, conservation of the $U(1)$ $R$-symmetry
    current requires scale invariance.
    Non-scale invariant theories will have anomalies in the $R$-symmetry
    current, although it is often then possible to combine the anomalous
    $R$-symmetry with an anomalous chiral symmetry to yield a non-anomalous
    $R$-symmetry.
    If there is no non-anomalous $R$-symmetry,
    then the $R$-charge density will not be a
    hydrodynamic degree of freedom
    For such theories, just ignore all mention of the $R$-symmetry
    current in the following discussion.
    }

Since the supercurrents are conserved, one might expect that
fluctuations in supercharge densities will be unable to relax locally,
just as ordinary current conservation directly leads to diffusive
relaxation of bosonic conserved charge densities.
And slow relaxation of supercharge densities may in turn generate
long-time tails in {\em bosonic} observables which can couple
quadratically to supercharge densities.
In other words, even though supercharge densities are fermionic operators
with vanishing expectation value in any conventional statistical ensemble,
the dynamics of supercharge fluctuations may produce distinctive effects
in ordinary bosonic observables.
We will see that these expectations are largely correct.

\subsection{Supercharge fluctuations}
\label{sec:susy thermodynamics}

The hydrodynamic nature of supercharge densities is
reflected in the long-time behavior of the corresponding
correlation function,
\begin{equation}
C^{\mu\nu}_{\dot{\alpha}\beta}(x)\equiv\langle \bar{S}^{\mu}_{\dot{\alpha}} (x)
 \, S^{\nu}_{\beta}(0)\rangle \ .
\end{equation}
Finite-temperature thermal ensembles are not invariant under
supersymmetry transformations.%
\footnote
    {%
    Formally, the transformation of a statistical density matrix $\rho$
    under any infinitesimal symmetry transformation is defined 
    by the condition that
    $
	\Tr (\delta \rho \> {\cal O} )
	=
	\Tr (\rho \> \delta {\cal O} )
    $,
    where $\cal O$ is any observable.
    This is a generalization of the relation between
    Schrodinger and Heisenberg pictures, and expresses
    the equivalence between active and passive
    views of symmetry transformations.
    If $\cal G$ is the generator of the transformation, then
    the infinitesimal transformation of the observable is
    $\delta {\cal O} = [ {\cal G}, {\cal O} ]$.
    For normal (bosonic) symmetries, this of course implies that
    $\delta \rho = -[ {\cal G}, \rho]$,
    so invariance under the symmetry means the density matrix $\rho$
    commutes with the symmetry generator $\cal G$.
    But for an infinitesimal supersymmetry transformation,
    ${\cal G} = \xi^\alpha Q_\alpha$
    with $\xi$ an external Grassmann spinor
    which anticommutes with all fermionic operators.
    In this case, one has
    $
	\Tr (\rho \, [\xi Q,{\cal O}] )
	=
	\Tr (\xi \, \rho \, [Q,{\cal O}] )
	=
	\Tr ([\xi \rho, Q] \, {\cal O} )
	=
	\Tr (\{\rho, \xi Q\} \, {\cal O})
    $,
    so the variation of the density matrix is the anticommutator with
    the supersymmetry generator,
    $
	\delta\rho = \{ \rho, \xi Q \} = \xi \, \{ \rho, Q \}
    $,
    not the commutator.
    [We assume throughout that $\rho$ and $\cal O$ are bosonic operators
    --- possibly constructed by multiplying odd numbers of fermionic operators
    by independent external Grassmann spinors.]
    Therefore, even though the equilibrium canonical ensemble
    $\rho = Z^{-1} \, e^{-\beta H}$ commutes with the supercharge $Q$,
    this does not mean that the canonical ensemble is invariant under
    supersymmetry.%
    \label {fn:hot susy}%
    }
This is reflected in the fact that at finite temperature bosons
and fermions obey different statistics (Bose-Einstein or Fermi-Dirac),
as well as in the fact that bosonic and fermionic fields obey
different boundary conditions (periodic versus anti-periodic)
in Euclidean functional integrals representing equilibrium thermal systems.
But it is important to understand that
the non-invariance of thermal ensembles under supersymmetry
does not imply that the supercurrent fails to be conserved.
The conservation of the supercurrent,
\begin {equation}
    \partial_\mu S^{\mu}_{\alpha}(x)=0
\end {equation}
is an operator identity, valid in any physical state.
Since thermal averages are just linear combinations of
expectation values in physical states,
the conservation of the supercurrent implies that
the thermal supercurrent correlation function satisfies the Ward identity%
\footnote
    {%
    This assumes one suitably defines the short-distance subtraction
    in the renormalization of the product of supercurrents to remove what
    would otherwise be contact terms on the right hand side.
    One can always do so, provided the supersymmetric theory
    under discussion exists.
    }
\begin{equation} \label{eq:canonWI}
  \partial_\mu \, C_{\dot\alpha \beta}^{\mu\nu}(x)=0 \,.
\end{equation}
Similarly, taking thermal expectations of both sides of
the supersymmetry algebra (\ref {eq:susyalg}) shows that
mean square fluctuations in supercharge are directly
related to the total energy of the thermal ensemble.
A spatial Fourier transform of the Ward identity (\ref {eq:canonWI}) gives
\begin {equation}
    \partial_0 \, \tilde C^{0\nu}_{\dot\alpha \beta}(\k,t)
    =
    -i k_j \, \tilde C^{j\nu}_{\dot\alpha \beta}(\k,t)
    =
    O(k) \,,
\end {equation}
which directly shows that
long wavelength fluctuations in supercharge density 
must relax arbitrarily slowly.

\subsection {Super-thermodynamics}

Having realized that supercharge densities are hydrodynamic degrees of freedom
in supersymmetric theories,
we would like to construct the appropriate hydrodynamic equations
in complete analogy with the earlier non-supersymmetric treatment.
To do so, we must construct the appropriate
constitutive relations for the spatial part of the supercurrent
(and the $R$-symmetry current),
and also understand possible new terms involving products of
superchange densities which might appear in the constitutive relations
for the stress tensor and other ordinary symmetry currents.

As discussed previously, non-derivative terms in constitutive relations
are completely determined by equilibrium thermodynamic derivatives
of the free energy.
This is true provided one has introduced
chemical potentials conjugate to all conserved densities of interest.
To do this for a supersymmetric theory, we must first understand
the generalization of ordinary thermodynamics to `equilibrium'
ensembles in which there is a non-zero expectation value of the supercharge,
produced by turning on a super-chemical potential.

To construct an equilibrium state with non-zero supercharge density,
consider generalizing the usual statistical operator (\ref{eq:density operator})
by adding supercharges multiplied by Grassmann-valued
super-chemical potentials,%
\footnote
    {%
    Though fermionic chemical potentials may seem peculiar,
    one can regard them as a non-dynamical
    gravitino field, to which the theory is coupled.
    It should be stressed that these generalized density operators
    do not have the usual statistical interpretation.
    Their diagonal matrix elements, although commuting,
    are not real numbers, and therefore cannot be
    viewed as statistical probabilities.
    Nevertheless, these generalized density operators may formally
    be regarded as equilibrium ensembles, in which generalized
    thermodynamic relations hold.
    The logical alternative of introducing
    real-valued chemical potentials for supercharges
    is formally inconvenient and also fails to
    yield any satisfactory physical interpretation.
    In particular, a density matrix containing a real-valued 
    chemical potential for supercharges,
    when applied to the union of two disjoint physical systems,
    does not factorize, even if the two systems are causally
    disconnected~\cite{real mu for supercharge}.
    }
\begin {equation}
\label{eq:susy density operator}
    \hat \rho_{\rm s}
    \equiv
    Z^{-1} \,
    \exp \left[
	\beta \, ( u_\nu P^\nu +
	\mu^\alpha Q_\alpha + \bar\mu_{\dot\alpha} \bar Q^{\dot\alpha} )
    \right] .
\end {equation}
To simplify expressions,
we are temporarily suppressing any chemical potentials for bosonic conserved
charges.
Note that the addition of spinorial chemical potentials coupled to
the supercharges $Q_\alpha$ and $\bar Q_{\dot\alpha}$ maintains
translation invariance of the equilibrium state
but explicitly breaks rotation invariance.

In the ensemble described by $\hat\rho_{\rm s}$,
the supercharges $Q_\alpha$ and $\bar Q_{\dot\alpha}$
have non-zero expectation values
proportional to $\bar\mu$ and $\mu$, respectively.
More precisely,
to first order in $\mu$ and $\bar\mu$,
\begin {eqnarray}
    \volume^{-1} \,
    \left\langle
    \xi^\alpha Q_\alpha
    \strut\right\rangle_{\mu,\bar\mu}
    &\equiv&
    \Tr \left(\hat \rho_s \> \xi^\alpha Q_\alpha \strut\right)
    \bigm/
    \volume
\nonumber \\ &=&
    -\beta
    \left\langle
	\bar\mu^{\dot\alpha} \bar Q_{\dot\alpha}
	\> \xi^\alpha Q_\alpha
    \right\rangle / \, \volume
\nonumber \\ &=&
    -\beta \,
    \xi^\alpha \bar\mu^{\dot\beta} 
    \left\langle \half \{ Q_\alpha, \bar Q_{\dot\beta} \} \right\rangle
    / \volume
\nonumber \\ &=&
    -\beta \,
    \xi^\alpha \,\sigma^\nu_{\alpha\dot\beta} \,\bar\mu^{\dot\beta}
    \left\langle T^0_{\;\;\nu} \right\rangle
\nonumber \\ &=&
    \beta \,
    \xi^\alpha
    \bigl(
	\bar\vareps \, \sigma^0
	- \bar\enthalpy \, {\bm \sigma} \cdot \u
    \bigr)_{\alpha\dot\beta}
    \> \bar\mu^{\dot\beta}
    \,,
\label {eq:xi Q}
\end {eqnarray}
where $O(\u^2)$ corrections have been dropped in the last equality.
Here $\xi$ is an independent Grassmann spinor.
The intermediate expectations in (\ref {eq:xi Q}) denote expectation
values in the usual canonical ensemble (with 4-velocity $u$),
and trace cyclicity and the supersymmentry algebra (\ref {eq:susyalg})
have been used.
Expanding to higher order in $\mu$ and $\bar\mu$ would lead to corrections
of order $\mu \bar\mu^2$ which will not be relevant for our purposes.
Similarly,
\begin {equation}
    \volume^{-1}
    \left\langle
    \bar\xi^{\dot\alpha} \bar Q_{\dot\alpha}
    \right\rangle_{\mu,\bar\mu}
    \equiv
    \volume^{-1} \,
    \Tr \left(\hat \rho_s \> \bar\xi^{\dot\alpha} \bar Q_{\dot\alpha} \right)
    =
    \beta \,
    \mu^\alpha
    \bigl(
	\bar\vareps \, \sigma^0
	- \bar\enthalpy \, {\bm \sigma} \cdot \u
    \bigr)_{\alpha\dot\beta}
    \> \bar\xi^{\dot\beta}
    \,,
\label {eq:xi Qbar}
\end {equation}
up to corrections of order $\mu^2 \bar\mu$ [and $O(\u^2)$].

Naively, one would expect the generalized partition function
\begin {equation}
    Z(\beta,u,\mu,\bar\mu)
    \equiv
    \Tr 
    \exp \left[
	\beta \, ( u_\nu P^\nu +
	\mu^\alpha Q_\alpha + \bar\mu_{\dot\alpha} \bar Q^{\dot\alpha} )
    \right]
\end {equation}
to be a generating function for expectation values of supercharges
and products of supercharges.
However, this is not true.
Although the two terms we have added to the exponential defining
$\hat\rho_s$ both commute with the total momentum $P^\nu$,
they do not commute with each other.
As a result, the derivative
$(\partial/\partial\mu) \ln Z(\beta,u,\mu,\bar\mu)$
does not equal $\beta\langle Q \rangle$.
In fact, the generalized partition function is completely independent of
the super-chemical potentials, and simply equals the usual (physical)
partition function,
\begin {equation}
    Z(\beta,u,\mu,\bar\mu) = Z(\beta,u,0,0) = \Tr \> e^{\beta u_\nu P^\nu} \,.
\end {equation}
To see this, first note that
\begin{equation}
   \Tr \left[ \mu^\alpha \> e^{\beta u_\nu P^\nu} \cdots \right]
   =
   \mu^\alpha \, \Tr \left[ (-1)^F \, e^{\beta u_\nu P^\nu} \cdots \right] ,
\end{equation}
and therefore
\begin{equation}
   \frac{\partial }{ \partial \mu^\alpha} \,
   \Tr \left[ e^{\beta u_\nu P^\nu} \cdots \right]
   =
   \Tr \left[ (-1)^F \right.
       \frac{\partial }{ \partial \mu^\alpha} \left. e^{\beta u_\nu P^\nu} \cdots
   \right] .
\end{equation}
Here $(-1)^F \equiv e^{2 \pi i J_z}$ is the operator which
multiplies all bosonic states by $+1$ and all fermionic states by $-1$.
This operator commutes with $P^\nu$ and anticommutes with the supercharges.
Next, factorize the exponentials in $\hat \rho_s$ using the
Baker-Campbell-Hausdorff formula,
\begin {eqnarray}
    e^{\beta ( u_\nu P^\nu + \mu^\alpha Q_\alpha
	+ \bar\mu_{\dot\alpha}\bar Q^{\dot\alpha} ) }
    &=&
    e^{\beta u_\nu P^\nu} \>
    e^{\beta \mu^\alpha Q_\alpha} \>
    e^{\beta \bar\mu_{\dot\alpha} \bar Q^{\dot\alpha} } \>
    e^{-\frac{1}{2}\beta^2
	[\mu^\alpha Q_\alpha, \, \bar\mu_{\dot\alpha} \bar Q^{\dot\alpha}]}
\nonumber \\ &=&
    e^{\beta \mu^\alpha Q_\alpha} \>
    e^{\beta \bar\mu_{\dot\alpha} \bar Q^{\dot\alpha} } \>
    e^{\beta P_\nu (u^\nu -\beta
	\mu^\alpha \, \sigma^\nu_{\alpha\dot\alpha} \bar\mu^{\dot\alpha} )}
    \,.
\end {eqnarray}
Consequently,
\begin {eqnarray}
    \frac{1 }{ \beta} \,
    \frac{\partial }{ \partial \mu^\gamma} \,
    Z(\beta,u,\mu,\bar\mu)
    &=&
    \frac{1 }{ \beta} \,
    \Tr \left[(-1)^F \right.
	\frac{\partial }{ \partial \mu^\gamma} \left.
	e^{\beta \mu^\alpha Q_\alpha} \>
	e^{\beta \bar\mu_{\dot\alpha} \bar Q^{\dot\alpha} } \>
	e^{\beta P_\nu(u^\nu-\beta \mu^\alpha \, \sigma^\nu_{\alpha\dot\alpha}
	    \bar\mu^{\dot\alpha}) } \>
    \right]
\nonumber \\ &=&
    \Tr \left[(-1)^F
	\left( Q_\gamma -
	\beta \sigma^\nu_{\gamma\dot\alpha} \,\bar\mu^{\dot\alpha} P_\nu \right)
	e^{\beta \mu^\alpha Q_\alpha} \>
	e^{\beta \bar\mu_{\dot\alpha} \bar Q^{\dot\alpha} } \>
	e^{\beta P_\nu(u^\nu-\beta \mu^\alpha \, \sigma^\nu_{\alpha\dot\alpha}
	    \bar\mu^{\dot\alpha}) } \>
	\right]
\nonumber \\ &=&
    \Tr \left[(-1)^F \>
	e^{\beta \mu^\alpha Q_\alpha} \>
	e^{\beta \bar\mu_{\dot\alpha} \bar Q^{\dot\alpha} } \>
	e^{\beta P_\nu(u^\nu-\beta \mu^\alpha \, \sigma^\nu_{\alpha\dot\alpha}
	    \bar\mu^{\dot\alpha}) } \>
	\left( -Q_\gamma -
	\beta \sigma^\nu_{\gamma\dot\alpha} \,\bar\mu^{\dot\alpha} P_\nu \right)
	\right]
\nonumber \\ &=&
    \Tr \left[
	(-1)^F
	\left( - Q_\gamma +
	\beta \sigma^\nu_{\gamma\dot\alpha} \,\bar\mu^{\dot\alpha} P_\nu \right)
	e^{\beta \mu^\alpha Q_\alpha} \,
	e^{\beta \bar\mu_{\dot\alpha} \bar Q^{\dot\alpha} } \,
	e^{\beta P_\nu(u^\nu-\beta \mu^\alpha \, \sigma^\nu_{\alpha\dot\alpha}
	    \bar\mu^{\dot\alpha}) } \>
	\right] \!.\qquad
\label {eq:dZ/dmu}
\end {eqnarray}
The penultimate step uses the anticommutation of
$Q$ with $(-1)^F$ and trace cyclicity to move $Q$ from the left to the right
end of the trace.
The final step uses the supersymmetry algebra to commute $Q$ through
the exponential of $\bar Q$:
\begin {equation}
    \left[Q_\gamma \, , 
    e^{\beta \bar\mu_{\dot\alpha} \bar Q^{\dot\alpha}}\right]
    =
    \beta \bar\mu^{\dot\gamma} \{Q_\gamma , \bar Q_{\dot\gamma} \} \,
    e^{\beta \bar\mu_{\dot\alpha} \bar Q^{\dot\alpha}}
    =
    2\beta \, \sigma^\nu_{\gamma\dot\gamma} \, \bar\mu^{\dot\gamma} \, P_\nu
    \, e^{\beta \bar\mu_{\dot\alpha} \bar Q^{\dot\alpha}} \,.
\end {equation}
Since the last line in (\ref {eq:dZ/dmu}) is minus the second line,
this shows that $(\partial /\partial \mu^\gamma) Z(\beta,u,\mu,\bar\mu) = 0$.
One may show that
$(\partial/\partial \bar\mu_{\dot\gamma}) Z(\beta,u,\mu,\bar\mu)=0$
in a completely similar fashion.

In other words, although adding fermionic chemical potentials to the
statistical operator induces non-zero expectations values for the
supercharge densities, it does not change the partition function at all.
This means that the ``thermodynamic pressure'', defined as
$(\beta \volume)^{-1} \ln Z$, is not affected by the
presence of non-zero equilibrium supercharge densities.
In fact, this is required if the
thermodynamic pressure is to agree with the
microscopic pressure, defined in terms of the expectation value
of the stress tensor,
$\pressure \equiv \frac{1}{3} \Tr \, (\hat{\rho}_s \, T_{ii})$.
To see that the microscopic pressure is also
$\mu$-independent, one may repeat the steps shown in Eq.~(\ref {eq:dZ/dmu})
when there is an additional insertion of the stress-energy tensor.
This yields
\begin {equation}
    \frac{1 }{ \beta} \, \frac{\partial }{ \partial \mu^\alpha} \>
    \langle T^{\lambda\rho} \rangle_{\mu,\bar\mu}
    =
    \half
    \left\langle
	(-1)^F \, \left[ Q_\alpha \,,\, T^{\lambda\rho} \right]
    \right\rangle_{\mu,\bar\mu} .
\label {eq:dT/dmu}
\end {equation}
But the supersymmetry transformation of the stress-energy tensor
involves spacetime derivatives of the supercurrent
(see, {\em e.g.}, \cite{Sohnius}),
\begin{equation}
  \left[Q_\alpha,T^{\mu\nu}\right] = -{\textstyle \frac{i}{2}} \left\{
  (\sigma^{\mu\rho})_\alpha^{\ \beta} \, \partial_{\rho} S^{\nu}_\beta +
  (\sigma^{\nu\rho})_\alpha^{\ \beta} \, \partial_{\rho} S^{\mu}_\beta
  \right\} \,,
\end{equation}
whose expectation (\ref {eq:dT/dmu})
vanishes in the translationally invariant equilibrium ensemble.

None of the above results are affected by the presence of
non-zero chemical potentials for ordinary conserved charges
which commute the the supercharges.
But if a non-zero chemical potential $\mu_\smallR$ for the $R$-charge
is present,
then the generalized partition function does acquire
dependence on the fermionic chemical potentials. 
The lowest such term is proportional to $\mu \bar\mu \mu_\smallR$;
this term is directly related to the equilibrium form of the
$R$-charge current shown below.

In a similar fashion, one may evaluate the generalized equilibrium
expectation values of the bosonic symmetry currents and the supercurrents.
Ordinary bosonic conserved currents commute with the supercharges,
and hence are unaffected by super-chemical potentials.
In the presence of non-zero ordinary chemical potentials,
\begin {equation}
    \left\langle
	J^\mu_a
    \right\rangle_{\mu,\bar\mu}
    =
    \bar n_a \, u^\mu
\end {equation}
as usual,
where $\bar n_a$ is the equilibrium density of the conserved charge $N_a$.
For the supercurrents one finds,
\begin{subequations}
\begin {eqnarray}
    \left\langle
	\xi^\alpha \, S^\lambda_\alpha
    \right\rangle_{\mu,\bar\mu}
    &=&
    -\beta \,
    \bar T^{\lambda}_{\ \, \nu} \,
    (\xi^\alpha \, \sigma^\nu_{\alpha\dot\beta} \, \bar \mu^{\dot\beta})
    + O(\mu \bar\mu^2) \,,
\\[2pt]
    \left\langle
	\bar\xi^{\dot\alpha} \, \bar S_{\dot\alpha\lambda}
    \right\rangle_{\mu,\bar\mu}
    &=&
    -\beta \,
    \bar T^{\lambda}_{\ \, \nu} \,
    (\mu^\alpha \, \sigma^\nu_{\alpha\dot\beta} \, \bar\xi^{\dot\beta})
    + O(\mu^2 \bar\mu) \,,
\end {eqnarray}
\label {eq:S equil}%
\end{subequations}
where
$\bar T^{\mu\nu} = \bar\enthalpy \, u^\mu u^\nu + \bar\pressure \, g^{\mu\nu}$
is the usual equilibrium form of the stress-energy tensor.
These results reflect the supersymmetry transformations
\begin{subequations}
\begin {eqnarray}
    \left\{ Q_\beta , \bar S^\lambda_{\dot\alpha} \right\}
    &=&
    2 \, T^\lambda_{\;\;\nu} \; \sigma^\nu_{\beta\dot\alpha} 
    + (\mbox{space-time gradients}) \,,
\\
    \left\{ \bar Q_{\dot\beta} , S^\lambda_\alpha \right\}
    &=&
    2 \, T^\lambda_{\;\;\nu} \; \sigma^\nu_{\alpha\dot\beta}
    + (\mbox{space-time gradients}) \,,
\end {eqnarray}
\label {eq:QS comm}%
\end{subequations}
and
$
    \{ Q_\beta, S^\lambda_\alpha \}
$,
$
    \{ \bar Q_{\dot\beta}, \bar S^\lambda_{\dot\alpha} \}
$
being pure space-time gradients.
(The expectation value of the space-time gradient terms vanish
in all translationally invariant equilibrium states.)
For the $\lambda=0$ components of these commutators,
the unspecified space-time gradients are strictly spatial gradients
which vanish on spatial integration, so that
the anti-commutation relations (\ref {eq:QS comm}) are consistent with the
supersymmetry algebra (\ref {eq:susyalg}).
For the $R$-symmetry current, one finds
\begin {equation}
    \left\langle
	J^\lambda_\smallR
    \right\rangle_{\mu,\bar\mu}
    =
    \bar n_\smallR \, u^\lambda
    - \half \, \beta^2 \, \bar T^{\lambda}_{\;\;\,\nu} \, 
    (\mu^\alpha \sigma^\nu_{\alpha\dot\beta} \, \bar \mu^{\dot\beta})
    + O(\mu^2 \bar\mu^2) \,,
\label {eq:JR}
\end {equation}
where $\bar n_\smallR$ is the equilibrium $R$-charge density in the
absence of super-chemical potentials.
The term quadratic in super-chemical potentials arises from the
non-trivial commutators of the supercharges with the $R$-current,%
\begin{subequations}
\begin {eqnarray}
    \left[ Q_\alpha, J^\mu_R \vphantom {\bar Q}\right]
    &=&
    S^\mu_\alpha +(\mbox{space-time gradients})\,,
\\
    \left[ \bar Q_{\dot\alpha}, J^\mu_R \right]
    &=&
    -\bar S^\mu_{\dot\alpha} +(\mbox{space-time gradients})\,.
\end {eqnarray}
\end{subequations}
Note that the relation (\ref {eq:JR})
implies that non-zero super-chemical potentials
induce a non-zero $R$-charge density even when the $R$-charge chemical
potential vanishes.

\subsection{Supersymmetric constitutive relations}

We can now construct the required constitutive relations for
the stress tensor and supercurrent. 
As before, we will need to include
all possible terms which are linear in fluctuations away from
a reference equilibrium state (chosen to have vanishing charge, supercharge,
and spatial momentum densities)
involving one spatial derivative, as well as
terms without derivatives which are linear or quadratic in fluctuations.
Neglected terms involving either more derivatives or more fluctuations
will not affect the leading long-time tails in stress-stress or
current-current correlators.

We will use $\rho_\alpha$ and $\bar\rho_{\dot\alpha}$ to denote
the (Grassmann valued) supercharge densities in the
non-equilibrium ensemble of interest. 
In equilibrium, the relations (\ref {eq:S equil}) imply that
the supercharge densities are related to the super-chemical potentials via
\begin {equation}
    \rho_\alpha
    =
    \beta \left(
	\bar\vareps \, \sigma^0 - \bar\enthalpy \, {\bm \sigma}\cdot\u
    \right)_{\alpha \dot\beta} \, \bar\mu^{\dot\beta}
    \,, \qquad
    \bar\rho_{\dot\alpha}
    =
    -\beta \, \mu^\beta
    \left(
	\bar\vareps \, \sigma^0 - \bar\enthalpy \, {\bm \sigma}\cdot\u
    \right)_{\beta\dot\alpha} \,,
\label {eq:rho}
\end {equation}
neglecting $O(\u^2)$ corrections.
For the reasons discussed in the previous subsection,
to quadratic order in deviations away from the reference equilibrium state,
the pressure contains no coupling to supercharge densities.
Consequently, the constitutive relation for the stress is identical
to the previous result (\ref{eq:Tij-std}),
except for the inclusion of $R$-charge density fluctuations along
with other bosonic charge fluctuations,%
\footnote{
    In writing the constitutive relation (\ref{eq:susyT}),
    we have assumed, for simplicity, that the
    global bosonic symmetry currents $\j_a$ are all vectors, rather
    than pseudo-vectors, and that parity is a symmetry of the theory.
    This prevents a $\,\Xi_{a\smallR} n_a n_\smallR$
    cross term from appearing in the constitutive relation (\ref {eq:susyT})
    for the stress,
    and likewise ensures that the $R$-charge density does not mix with
    other global charge densities in the constitutive relations
    (\ref{eq:susyJ}) and (\ref{eq:susyR}) for the currents.
    }
\begin{eqnarray}
\label{eq:susyT}
  \T^{ij} & = & \delta^{ij}
		\left[
		      \bar\pressure + \vs^2 \, \delta\epsilon +
                       \half \, \xi \, (\delta\epsilon)^2 +
                       \half \, \Xi_{ab} \, n_a n_b +
                       \half \, \Xi_\smallR \, n_{\!\smallR}^{\ 2} -
                       \vs^2 \, \frac{\bpi^2}{\bar\enthalpy}
		\right]
		+ \frac{\pi^i \pi^j}{\bar\enthalpy}
\nonumber \\ && \qquad {}
	       - \gamma_\zeta \, \delta^{ij} \, \grad \cdot \bpi
	       -\gamma_\eta \left( \nabla^i \pi^j + \nabla^j \pi^i
	       - {\textstyle \frac{2}{3}}\delta^{ij} \, \grad\cdot\bpi
	       \right) .
\end{eqnarray}

The constitutive relations for bosonic internal symmetry currents
are unchanged from the non-supersymmetric case,
\begin{equation} \label{eq:susyJ}
   \j_a = -D_{ab}\grad n_b +
   \frac{n_a \,\bpi}{\bar\enthalpy} \,,
\end{equation}
while the new constitutive relation for the $R$-symmetry current is
\begin{equation} \label{eq:susyR}
   \j_{\smallR}= -D_{\!\smallR} \grad n_\smallR +
                   \frac{n_\smallR \,\bpi}{\bar\enthalpy}
		   -
                   \frac{\bar\pressure}{2\bar\vareps^{\, 2}} \>
		   \bar\rho_{\dot\alpha}
		   \,\bar\sigmav^{\dot\alpha \beta}\,
                   \rho_\beta \,.
\end{equation}
The quadratic coupling to supercharge densities follows directly from
the (generalized) equilibrium form (\ref {eq:JR}) for the expectation
of the $R$-current, combined with the relation (\ref {eq:rho}).

The remaining constitutive relations are those of the supercurrents.
The appropriate form is more complicated than for bosonic currents
because the supercharge density is a spinor.
There are two independent rotationally covariant forms in which
a spatial gradient can be applied to a spinor $\rho_\alpha$ to form
a vector-spinor, namely $\nabla^i \rho_\alpha$ and
$(\sigma^{ij})_\alpha^{\;\;\beta} \, \nabla_{\!j} \, \rho_\beta$.
Consequently, the supercurrent constitutive relations have the form
\begin{subequations}
\begin{eqnarray}
    S_\alpha^i &=&
    - D_s  \nabla^i \rho_\alpha
    - D_\sigma (\sigma^{ij})_{\alpha}^{\;\;\beta} \, \nabla_{\!j} \, \rho_\beta
    -\frac{\bar\pressure}{\bar\vareps} \,
    (\sigma^i \bar{\sigma}^0)_{\alpha}^{\;\;\beta} \, \rho_\beta
    + \frac{1}{\bar\vareps} \, \rho_\alpha \,\pi^i
    + \frac{\bar\pressure}{\bar\vareps^2}\; \pi^k
	(\sigma^i \bar\sigma^k)_{\alpha}^{\;\;\beta} \, \rho_\beta
     \,,\quad
\\[3pt]
    \bar S_{\dot\alpha}^i &=&
    - D_s  \nabla^i \bar\rho_{\dot\alpha}
    - D_\sigma \nabla_{\!j} \, \bar\rho_{\dot\beta} \,
      (\bar\sigma^{ji})^{\dot\beta}_{\;\;\dot\alpha} 
    -\frac{\bar\pressure}{\bar\vareps} \,
    \bar\rho_{\dot\beta} \,
    (\bar\sigma^0 \sigma^i)^{\dot\beta}_{\;\;\dot\alpha} \,
    + \frac{1}{\bar\vareps} \, \bar\rho_{\dot\alpha} \,\pi^i
    + \frac{\bar\pressure} {\bar\vareps^2}\; \pi^k
	\bar\rho_{\dot\beta} \,
    (\bar\sigma^k \sigma^i)^{\dot\beta}_{\;\;\dot\alpha} \,
    \,.\quad
\end{eqnarray}
\label{eq:susyS}%
\end{subequations}
Here $D_s$, $D_\sigma$ are two new diffusion constants,
which multiply the two structures allowed by rotation invariance.%
\footnote
    {
    Assuming parity invariance of the theory ensures that
    the diffusion constants for the left and right handed supercurrents
    are the same.
    The Kubo formula for the supercharge diffusion constant $D_s$ is
    $
	\bar\vareps \, D_s = -\lim_{\omega\to 0}
	\frac{1}{12}\int d^4 x \, e^{i\omega x^0}
	\bar\sigma^{0\dot\alpha\beta}
	\, C^{ii}_{\dot\alpha\beta}(x)
    $.
    }
The remaining non-derivative terms are dictated by the equilibrium
results (\ref {eq:S equil}), when expanded to first order in flow velocity.
The third term, which is linear in supercharge density, reflects the fact
that a constant supercharge density is not rotationally invariant, and hence
induces a non-zero constant spatial supercurrent.
The convective terms are more complicated
than just $\rho_\alpha \pi^i/\bar\enthalpy$,
because of the presence of this
linear non-derivative term.
If the supersymmetric theory in question is also
scale invariant, then  $\bar\sigma^\mu S_\mu$ must vanish,
as it is in the same supersymmetry multiplet as the trace
of the energy-momentum tensor.
Applying this constraint to
the constitutive relation (\ref{eq:susyS}),
one finds that the two diffusion constants must coincide,
$D_\sigma = D_s$,
in scale-invariant supersymmetric theories.

\subsection {Relaxation of supercharge fluctuations}

The constitutive relation (\ref{eq:susyS}),
combined with conservation of the supercurrent,
allows one to determine how long-wavelength fluctuations in
supercharge density relax.
Inserting the constitutive relation into the continuity equation
$\partial_t \, \rho_\alpha = -\grad \!\cdot\! \mbox{\boldmath $S$}_\alpha$,
neglecting the non-linear convective terms, and Fourier transforming in space
yields
\begin {equation}
    \left[
	\strut(\partial_t + D_s \, \k^2) \, \delta_\alpha^\beta
	- i \cs \, (\k \cdot \sigmav \, \bar{\sigma}^0)_\alpha^{\ \beta}
    \right]
    \rho_\beta(t,\k)
    = 0 \,,
\end {equation}
where $\cs \equiv \bar\pressure/ \bar\vareps$.
The solution in the small-$\k$ limit shows weakly damped, propagating behavior,
\begin{equation} \label{eq:rho-evolution}
   \rho_\alpha(t,\k)
   =
   e^{-D_s \, \k^2 t}
    \left[
	\delta_\alpha^\beta \cos(|\k| \cs t)
	+ i (\hat{\k} \cdot \sigmav \bar{\sigma}^0)_\alpha^{\;\;\beta} \,
	     \sin(|\k| \cs t)
    \right]
    \rho_\beta(0,\k) \,.
\end{equation}
This is a propagating collective excitation which is distinct from
normal sound.
In some respects, it is analogous to second sound in a superfluid.
Note that the velocity $\cs = \bar\pressure/\bar\vareps$
of these ``supersound waves'' is always less than
the ordinary speed of sound
$\vs = \sqrt {\partial \bar\pressure/\partial \bar\vareps}$.
In scale invariant theories, $\cs = 1/3$ while $\vs = 1/\sqrt 3$.
Collective excitations in supercharge density were discussed earlier in
Ref.~\cite{supersound}, from a somewhat different perspective.

Just as in the previous non-supersymmetric analysis,
the linearized non-equilibrium relaxation (\ref {eq:rho-evolution})
may be converted into a statement about the time-dependence of the
equilibrium supercharge density correlator.
Let
\begin {equation}
    C(t,\k)_{\alpha\dot\beta}
    \equiv
    \int d^3\x \> e^{-i \k\cdot\x} \>
    \bigl\langle
	S^0_\alpha(t,\x) \, \bar S^0_{\dot\beta}(0,{\bf 0})
    \bigr\rangle \,,
\end {equation}
where the expectation value is in the reference equilibrium state
with vanishing chemical potentials.
The small $\k$ limit of the equal time correlator
is fixed by the supersymmetry algebra,%
\footnote{
    In our conventions, $\sigma^0=-1$, so $C(t{=}0,\k{=}0)$
    is strictly positive,
    as the equal-time correlation function of any
    operator with its Hermitian conjugate must be.}
\begin {equation}
    C(0,\k{=}0)_{\alpha\dot\beta}
    =
    -\bar\vareps  \, \sigma^0_{\alpha\dot\beta} \,.
\end {equation}
Consequently, the linear response result (\ref {eq:rho-evolution})
is equivalent to the statement that, in the small $\k$ limit,
the equilibrium supercharge correlator for large times has the form
\begin{eqnarray}
  C(t,\k)_{\alpha\dot\beta}
  =
  - \bar\vareps \, e^{-D_s \, \k^2 |t|}
  \left[
      \sigma^0_{\alpha\dot\beta} \, \cos(|\k| \cs t)
      + i (\hat\k \cdot \sigmav)_{\alpha\dot\beta} \, \sin(|\k| \cs t)
  \right] .
\label{eq:supercharge-corr-fn-evolution}
\end{eqnarray}
One finds the same result for the small $\k$ behavior of the
supercharge correlator with the opposite  ordering,
\begin{eqnarray}
    \bar C(t,\k)_{\alpha\dot\beta}
    &\equiv&
    \int d^3\x \> e^{-i \k\cdot\x} \>
    \bigl\langle
	\bar S^0_{\dot\beta}(t,\x) \, S^0_\alpha(0,{\bf 0})
    \bigr\rangle
\nonumber\\ &=&
  - \bar\vareps \, e^{-D_s \, \k^2 |t|}
  \left[
      \sigma^0_{\alpha\dot\beta} \, \cos(|\k| \cs t)
      + i (\hat\k \cdot \sigmav)_{\alpha\dot\beta} \, \sin(|\k| \cs t)
  \right] .
\end{eqnarray}

\subsection {Supersymmetric long-time tails}

As in the non-supersymmetric case, second order
terms in the constitutive relations are sufficient to understand
the leading long-time tails in $\k{=}0$ correlators of all conserved currents.
Proceeding in exactly the same way as in
Section~\ref{sec:stress tails},
one sees from the absence of any $\bar\rho \rho$ term
in the stress tensor constitutive relation (\ref {eq:susyT})
that supercharge density fluctuations
have no effect on the leading long-time tail in the stress-stress correlator.
Consequently, the stress-stress correlator
has the same form as in the non-supersymmetric case
[{\em i.e.}, Eq.~(\ref{eq:stresstail-conformal}) for scale-invariant theories,
or Eq.~(\ref{eq:stresstail}) for the general case]
provided the $R$-charge is included as one of the global bosonic charges.
Similarly, the leading long-time tails (\ref{eq:JJtail}) in correlators
of ordinary bosonic symmetry currents
are unaffected by the presence of supercharge density fluctuations.

In contrast,
the presence of a quadratic supercharge density term in the
constitutive relation (\ref {eq:susyR}) for the $R$-symmetry current
means that the long time tail in the $R$-symmetry current correlator
is affected by supercharge fluctuations.
One finds that the leading large time behavior is
\begin{equation}
  \int d^3\x \>
  \left\langle
      \half \{ j_\smallR^k(t,\x), j_\smallR^{\,l}(0,{\bf 0}) \}
  \right\rangle
  \sim
  \frac{\delta^{kl} }{ 12 }
  \left\{
      \frac{(T/\bar\enthalpy) \, \chi_\smallR}
	  {[(D_{\smallR} {+} \gamma_\eta) \, \pi |t|]^{3/2}}
      + \frac{\coeff 14 c_s^2  }{ [2D_{s} \, \pi |t|]^{3/2}}
  \right\} .
\end{equation}

If the same analysis is applied to the
spatial supercurrent correlation function,
one finds that the second order $\pi \rho$ terms in
the constitutive relation (\ref{eq:susyS}) do not generate long-time tails.
Because the speed of ordinary sound ($\vs$)
and supersound ($\cs$) differ,
fluctuations (with the same wavevector)
in momentum density and supercharge density
cannot remain in phase.
This is analogous to the fact that only transverse momentum
density fluctuations, not longitudinal sound waves,
contribute to the long-time tail (\ref {eq:JJtail}) of
ordinary current-current correlators.
So the supercurrent correlator
$\langle \bar{S}^i_{\dot\alpha}(t) S^j_\alpha(0) \rangle$,
at $\k{=}0$,
should decay exponentially in time.%
\footnote
    {%
    Because $\cs < \vs$, one finds that higher-order $\rho \pi^n$
    terms in the supercurrent constitutive relation
    also do not generate (higher-order) power-law tails.
    }

The differing long-time behaviors of the stress tensor
and supercurrent correlators does not
conflict with the fact that the stress-energy tensor
and supercurrents belong to the same supersymmetry multiplet.
The supersymmetry algebra implies simple
relations between $(-1)^F$ inserted thermal correlation
functions of conserved currents, but not between
conventional thermal correlation functions, for the reasons
noted in footnote \ref {fn:hot susy}.

\section{Long-time tails in $\mathcal{N}=4$ supersymmetric
Yang-Mills theory}
\label{sec:application to SYM}

As a specific application, the above discussion (trivially generalized)
may be applied to $\mathcal{N}{=}4$ supersymmetric $SU(\Nc)$ Yang-Mills theory,
at the origin of the moduli space and at non-zero temperature.%
\footnote
    {%
    The vacuum degeneracy associated with a non-trivial moduli space
    is lifted at non-zero temperature.
    For sufficiently high temperatures, the only stable equilibrium
    states lie at the origin of moduli space.
    }
Once again,
hydrodynamic degrees of freedom are densities of conserved charges.
In $\mathcal{N}{=}4$ supersymmetric Yang-Mills theory,
the supersymmetry algebra%
\footnote
    {%
    See, for example, Ref.~\cite{Sohnius}, for an introduction
    to the $\mathcal{N}{=}4$ algebra.
    }
is generated by the charges of
the dilation current
$
    j^{\mu}_D=x_\nu \T^{\mu\nu}
$,
the special conformal currents
$
    K^\mu_{\;\;\lambda}
    =
    2 x_\lambda x^\nu \T^\mu_{\;\;\nu} - x^2 \T^\mu_{\;\;\lambda}
$,
and the special supersymmetry currents
$
    \mathcal{S}^\mu_{A \alpha}
    =
    x_\nu \, \sigma^\nu_{\alpha\dot{\alpha}} \, \bar{S}^{\mu\dot{\alpha}}_A
$
and
$
    \bar{\mathcal S}^\mu_{A \dot\alpha}
    =
    x_\nu \, \bar{\sigma}^\nu_{\dot\alpha{\alpha}} \, S^{\mu\alpha}_A\,
$,
in addition to the stress-energy tensor $T^{\mu\nu}$ and supercurrents
$S^\mu_{A \alpha}$, $\bar S^\mu_{A \dot\alpha}$.
The index $A = 1, ..., 4$ is the extended supersymmetry label.
There are also 15 bosonic $R$-symmetry currents, whose charges
generate a non-anomalous global $SU(4)_R$ symmetry.

The currents 
$j^\mu_D$, $K^\mu_{\;\;\lambda}$, ${\mathcal S}^{\mu}_{A\alpha}$,
and $\bar{\mathcal S}^\mu_{A\dot\alpha}$
are all constructed from the energy-momentum tensor and ordinary supercurrents.
Therefore,
the time evolution of the charge densities of these currents is completely
determined by the time evolution of the energy-momentum tensor
and the supercurrent.
In other words, $j^0_D$, $K^0_{\;\;\lambda}$,
${\mathcal S}^0_{A\alpha}$, and $\bar{\mathcal S}^0_{A\dot\alpha}$
are not new hydrodynamic degrees of freedom,
independent from $\T^{0\nu}$, $S^{0}_{A\alpha}$, and $\bar{S}^0_{A\dot\alpha}$.
No Abelian magnetic fields appear as hydrodynamic degrees
of freedom, since the gauge group of the theory is $SU(\Nc)$, not $U(\Nc)$.
Moreover, holographic arguments imply that
the theory at non-zero temperature
has a mass gap \cite{Witten2},
in accord with the assumption
of finite correlation length made in Section~\ref{sec:general HD}.

The discussion of supercharge fluctuations in the previous
section generalizes easily to the case of $\mathcal{N}{=}4$ supersymmetry.
In particular, the constitutive relation
for the stress tensor has the same form (\ref{eq:susyT}),
except for the addition of a fundamental $SU(4)_R$ index
on the supercharges $\rho_{A\alpha}$
and an adjoint index on the $R$-charge densities $n_{\smallR \, a}$
({\em i.e.}, $a = 1, ... , 15$).
The theory is scale-invariant, and
therefore $\bar\varepsilon=3\bar\pressure$,
and the bulk viscosity vanishes,
as does the coefficient $\Xi_R$.
Hence,
fluctuations in the $R$-charge densities do not
contribute to constitutive relation for stress to second order,
which therefore must have the same form (\ref{eq:Tij scale-inv})
as in a non-supersymmetric scale-invariant theory.
Consequently, the long-time tail for the stress tensor correlator
(at $\k{=}0$) is given by the previous result (\ref{eq:stresstail-conformal}),
and the equivalent small frequency form of the spectral density is given by
Eq.~(\ref{eq:Tij spectral density scale-inv}).

The constitutive relation for supercurrent
(\ref{eq:susyS}) stays unchanged to second order
(except for the addition of the index $A$, and the specialization
to $D_s = D_\sigma$ and $\bar\pressure/\bar\vareps = 1/3$).
No long-time tail is present in the supercurrent correlator,
for the reasons discussed in the previous section.
Hence, the corresponding spectral density is analytic at
zero frequency.

The constitutive relations for the $R$-symmetry currents become
\begin{equation}
\label{eq:susy4R}
    \j_{\smallR \,a} =
              -D_\smallR \grad n_{\!\smallR \,a}
              + \frac{n_{\smallR \,a} \, \bpi }{ \bar\enthalpy}
              - \frac{1}{3\bar\vareps} \>
              \bar\rho_{A \dot\alpha} \, (t_a)^{AB} \,
	      \bar\sigmav^{\dot\alpha \beta}
              \, \rho_{B \beta} \ ,
\end{equation}
%
where $t_a$ are the generators of $SU(4)$, normalized so that
$\mathrm{tr}(t_a t_b) = \frac{1}{2}\delta_{ab}$.
The resulting long-time tails in the
$R$-symmetry current correlators (at $\k{=}0$) are
\begin{equation}
\label{eq:SYM R-tails}
   \int d^3\x \;
   \langle
       \half \{j_{\!\smallR \, a}^k(t,\x), j_{\!\smallR \, b}^l(0,{\bf 0}) \}
   \rangle
   \sim
   \frac{\chi_\smallR T }{ 16 \, \bar\vareps} \>
  \frac{\delta^{kl} \, \delta_{ab} \
   }{ [(D_{\! \smallR} {+} \gamma_\eta) \, \pi |t|]^{3/2}} +
  \> \frac{1 }{ 216 } \,
  \frac{\delta^{kl} \, \delta_{ab}  }{ [2D_{s} \, \pi |t|]^{3/2}} \,.
\end{equation}
The equivalent small-frequency expansion of the spectral density
is given by
\begin{equation}
\label{eq:JJ spectral density SYM}
     \frac{1 }{ \omega}
     \int dt\, d^3\x \ e^{i\omega t}
     \langle
     [j_{\!\smallR \, a}^k(t,\x), j_{\!\smallR \, b}^l(0,{\bf 0})]
     \rangle =
     \delta_{ab} \, \delta^{kl} \left( c^{(0)} +
     c^{(1)} \sqrt{|\omega|} \right) +
      O(|\omega|^{3/2}) \ ,
\end{equation}
where
\begin {subequations}
\begin{eqnarray}
c^{(0)} &=& \frac{2D_{\!\smallR} \, \chi_{\!\smallR}}{T} \ ,
\\
  c^{(1)} &=& - \frac{\chi_{\! \smallR}}{2\pi \bar\varepsilon} \,
              [2(D_\smallR {+} \gamma_\eta)]^{-3/2} -
              \frac{1}{216\pi T} \, D_s^{-3/2} \,.
\end{eqnarray}%
\end {subequations}
In writing expressions (\ref{eq:susy4R}) and
(\ref{eq:SYM R-tails})
we have used the fact that in a thermal state
without chemical potentials for the $SU(4)_R$ charges,
the matrix of diffusion constants, as well as the $R$ charge
susceptibility, must be invariants of the group:
$(D_{\!\smallR})_{ab}=D_{\!\smallR} \, \delta_{ab}$,
$(\chi_{\! \smallR})_{ab} = \chi_{\! \smallR} \, \delta_{ab}$.

The amplitudes of the long-time tails for the stress tensor correlator
(\ref{eq:stresstail-conformal}),
and the $R$-current correlator (\ref{eq:SYM R-tails}),
depend on the values of the
equilibrium energy density $\bar\vareps$,
the $R$-charge susceptibility $\chi_\smallR$,
the shear viscosity $\eta$, and $R$-charge diffusion constant $D_{\!\smallR}$.
These parameters must be
treated as input from short-distance physics.
In a weakly coupled theory,
these quantities can be computed in perturbation theory,
but no field-theoretic calculation is available
when the coupling constant is large.
However, the AdS/CFT correspondence predicts specific
values for these parameters in $\mathcal{N}{=}4$
supersymmetric Yang-Mills theory in the
limit of large $\Nc$ and large 't~Hooft coupling.
The equilibrium
energy density is evaluated from the thermodynamics of the dual
Anti-de~Sitter black hole, and is
predicted to be
\begin{equation}
  \label{eq:dual energy density}
  \bar\vareps = \frac{3\pi^2}{8} \, N_c^2 \, T^4 \,.
\end{equation}
The shear viscosity is evaluated as the zero-frequency limit of
the absorption cross-section of gravitons by the black three-brane
(with gravitons polarized parallel to the brane) \cite{Son et al},
and is predicted to be
\begin{equation}
  \label{eq:dual viscosity}
  \eta = \frac{\pi}{8} \, N_c^2 \, T^3 \,.
\end{equation}
Hence, in this limit
the ratio $\gamma_\eta \equiv \eta/\bar\enthalpy$
is predicted to equal $1/(4\pi T)$.
The diffusion constant and susceptibility for $SU(4)_R$ charges are extracted from the
pole structure of the thermal correlation function of $SU(4)_R$
charge densities \cite{AdSCFT diffusion}, and are predicted to be%
\begin{eqnarray}
  \label{eq:dual diffusion constant}
  D_{\!\smallR}&=&\frac{1}{2\pi T} \,, \\
\noalign{\hbox{and}}
  \chi_\smallR &=& \frac 18 \, N_c^2 \, T^3 \,.
\end{eqnarray}
The supercharge diffusion constant $D_s$ has, to our knowledge,
not yet been evaluated,
but should be computable by the methods of Ref.~\cite{AdSCFT diffusion}.

Although these strong coupling (and large $\Nc$) limits 
of transport coefficients, susceptibility, and energy density
cannot be compared with direct field-theoretic calculations,
they may be used as input into the effective hydrodynamic theory
(whose validity is independent of coupling and $\Nc$).
Inserting the AdS/CFT predictions
(\ref{eq:dual energy density}--\ref{eq:dual viscosity})
into the result (\ref{eq:Tij spectral density scale-inv})
yields the completely explicit form
\begin{eqnarray}
  \label{eq:predicted TT}
  \frac{1}{\omega} \int dx^0 \> d^3\x \; e^{i\omega x^0}
  \left\langle [\T_{ij}(x), \T_{kl}(0)] \right\rangle
    &=&
    \left(
	\half \, \delta_{ik} \delta_{jl}
	+ \half \, \delta_{il} \delta_{jk}
	- \coeff 13 \, \delta_{ij} \, \delta_{kl}
    \right) T^3
\nonumber\\ && {} \times
    \left[
	\frac{\pi}{2} \, N_c^2 -
	\frac{56 + 6^{3/2}}{60} \,
	\sqrt{\frac{\pi |\omega|}{T}}+
	O(|\omega|^{3/2})
    \right]
\end{eqnarray}
for the spectral density of the stress-stress correlator
in the strong coupling, large $\Nc$ limit.
This correlator is a non-trivial probe of the real-time dynamics
of strongly-coupled supersymmetric Yang-Mills plasma.
Because the AdS/CFT prediction of the supercharge diffusion constant
$D_s\,$ is currently lacking, we cannot give an equally explicit result for the
$R$-symmetry current spectral density (\ref{eq:JJ spectral density SYM}).

The result (\ref {eq:predicted TT}) displays the expected $1/\Nc^2$
relative suppression of the non-analytic part,
and also shows that the ratio of the leading analytic to the
leading non-analytic piece does not depend on
the coupling constant $g_{Y\!M}$ of the gauge
theory in the strong coupling limit.
In the language of the AdS/CFT correspondence,
$\alpha'/R_{AdS}^2=1/\sqrt{g^2_{Y\!M}\Nc}\;$, and $4\pi g_s = g^2_{Y\!M}$.
The ten-dimensional gravitational constant $\kappa$ scales as
$g_s \, \alpha'\,^2 \sim  R_{AdS}^4/\Nc$.
Thus, it is string/supergravity loops,
proportional to $\kappa^2$, which are responsible for $1/\Nc^2$
corrections in field theory correlators. The corresponding
one-loop amplitudes
are not straightforward to evaluate, given that even the scalar propagator
on the plane-symmetric AdS black hole background is not known analytically.%
\footnote{
    For progress in this direction,
    see Ref.~\cite{Starinets QN modes}.}
However, it would be quite interesting to reproduce hydrodynamic
results like Eq.~(\ref{eq:predicted TT}) directly from supergravity
loop corrections.
Just verifying the form of the small-frequency non-analyticity
from the gravity amplitudes,
without computing the precise coefficient, 
would be a worthwhile goal.

Finally, given that
in various AdS/CFT-like scenarios, finite-temperature field
theories are believed to be exactly equivalent to string
theories on backgrounds with thermal horizons,
it is natural to expect that there must exist a
correspondence between appropriate effective theories as well.
Namely, string theories on thermal
backgrounds should have low-energy descriptions which are dual
to effective long distance descriptions of finite-temperature field
theories, specifically hydrodynamics and kinetic theory
(in weakly coupled regimes).
As we have seen, constructing a
string theory dual of hydrodynamics will necessarily require
incorporating string loop effects in order to reproduce
aspects of long distance dynamics, such as long-time tails,
which are sensitive to the non-linearities of hydrodynamics.


\acknowledgments
\noindent

We thank Andreas Karch, Guy Moore, Dam Son, and Andrei Starinets
for numerous helpful conservations.
This work was supported, in part, by the U.S. Department
of Energy under Grant No.~DE-FG03-96ER40956.

\appendix
\section
    {Long-time tail in stress-stress correlator}
\label {app:non-scale-inv}

In a theory, which is not necessarily scale-invariant,
inserting the stress tensor constitutive relation (\ref {eq:Tij-std})
into the zero-wavenumber connected stress-stress correlator
gives
\begin {eqnarray}
    \volume^{-1}
    \left\langle \half \{ T^{ij}(t), T^{kl}(0) \} \right\rangle_{\rm conn}
    &=&
    \delta^{ij} \, \delta^{kl} \, \vs^2 \, \bar\enthalpy \, T
    +
    \frac{1 }{ \bar\enthalpy^2} \,
    H^{ij}_{mn} \, H^{kl}_{pq} \, B_{\pi^m\pi^n,\pi^p\pi^q}(t)
\nonumber \\ && {}
    +
    \frac{\xi }{ 2 \bar\enthalpy} \,
    \delta^{ij} \, H^{kl}_{pq} \, B_{\vareps\vareps,\pi^p\pi^q}(t)
    +
    \frac{\xi }{ 2 \bar\enthalpy} \,
    H^{ij}_{mn} \, \delta^{kl} \, B_{\pi^m\pi^n,\vareps\vareps}(t)
\nonumber \\ && {}
    +
    \frac{\xi^2 }{ 4} \,
    \delta^{ij} \, \delta^{kl} \, B_{\vareps\vareps,\vareps\vareps}(t)
    +
    \frac{\Xi_{ab} \, \Xi_{cd} }{ 4} \,
    \delta^{ij} \, \delta^{kl} \, B_{\rho_a\rho_b,\rho_c\rho_d}(t)
    \,,
\end {eqnarray}
where
$
    H^{ij}_{mn} \equiv
    \half \, \delta^i_m \, \delta^j_n
    + \half \, \delta^i_n \, \delta^j_m
    - \vs^2 \, \delta^{ij} \, \delta_{mn}
$,
and
\begin {eqnarray}
    B_{\pi^i\pi^j,\pi^k\pi^l}(t)
    &\equiv&
    \int d^3\x \>
    \left\langle
	\half \left\{
	    \pi^i(t,\x)\pi^j(t,\x), \,
	    \pi^k(0,{\bf 0})\pi^l(0,{\bf 0})
	\right\}
    \right\rangle_{\rm conn} \,,
\\
    B_{\pi^i\pi^j,\vareps\vareps}(t)
    &\equiv&
    \int d^3\x \>
    \left\langle
	\half \left\{
	    \pi^i(t,\x) \pi^j(t,\x), \, \vareps(0,{\bf 0})^2
	\right\}
    \right\rangle_{\rm conn} \,,
\\
    B_{\vareps\vareps,\pi^i\pi^j}(t)
    &\equiv&
    \int d^3\x \>
    \left\langle
	\half \left\{
	    \vareps(t,\x)^2, \, \pi^i(0,{\bf 0})\pi^j(0,{\bf 0})
	\right\}
    \right\rangle_{\rm conn} \,,
\\
    B_{\vareps\vareps,\vareps\vareps}(t)
    &\equiv&
    \int d^3\x \>
    \left\langle
	\half \left\{
	    \vareps(t,\x)^2, \, \vareps(0,{\bf 0})^2
	\right\}
    \right\rangle_{\rm conn} \,,
\\
    B_{\rho_a\rho_b,\rho_c\rho_d}(t)
    &\equiv&
    \int d^3\x \>
    \left\langle
	\half \left\{
	    \rho_a(t,\x) \, \rho_b(t,\x), \,
	    \rho_c(0,{\bf 0}) \, \rho_d(0,{\bf 0})
	\right\}
    \right\rangle_{\rm conn} \,.
\end {eqnarray}
Evaluating these correlators in the hydrodynamic regime,
as described in section \ref {sec:stress tails}, yields
\begin {eqnarray}
    B_{\pi^i\pi^j,\pi^k\pi^l}(t)
    &=&
    \int \frac{d^3\k }{ (2\pi)^3} \>
    \left[ \vphantom{\half}
	A_{\pi^i\pi^k}(t,\k) \, A_{\pi^j\pi^l}(t,-\k)
	+
	A_{\pi^i\pi^l}(t,\k) \, A_{\pi^j\pi^k}(t,-\k)
    \right]
\nonumber\\ &\sim&
    \frac{\bar\enthalpy^2 \, T^2 }{ 15}
    \left[
	\frac{
	    \delta^{ij} \delta^{kl} +
	    \delta^{ik} \delta^{jl} +
	    \delta^{il} \delta^{jk}
	}{
	    (4 \pi \gamma_s |t|)^{3/2}
	}
	+
	\frac{
	    2 \, \delta^{ij} \delta^{kl} +
	    7 \, ( \delta^{ik} \delta^{jl} + \delta^{il} \delta^{jk} )
	}{
	     \, (8 \pi \gamma_\eta |t|)^{3/2}
	}
    \right] ,
\\[2mm]
    B_{\pi^i\pi^j,\vareps\vareps}(t)
    &=&
    B_{\vareps\vareps,\pi^i\pi^j}(t)
    =
    \int \frac{d^3\k }{ (2\pi)^3} \;\;
	2 \, A_{\pi^i\vareps}(t,\k) \, A_{\pi^j\vareps}(t,-\k)
\nonumber\\ &\sim&
    \frac{\bar\enthalpy^2 \, T^2 }{ 3 \, v_s^2} \,
    \frac{\delta^{ij} }{ (4 \pi \gamma_s |t|)^{3/2}} \,,
\\[2mm]
    B_{\vareps\vareps,\vareps\vareps}(t)
    &=&
    \int \frac{d^3\k }{ (2\pi)^3} \;\;
	2 \, A_{\vareps\vareps}(t,\k) \, A_{\vareps\vareps}(t,-\k)
\nonumber\\ &\sim&
    \frac{\bar\enthalpy^2 \, T^2 }{ v_s^4} \,
    \frac{1 }{ (4 \pi \gamma_s |t|)^{3/2}} \,,
\\[2mm]
    B_{\rho_a\rho_b,\rho_c\rho_d}(t)
    &=&
    \int \frac{d^3\k }{ (2\pi)^3} \>
    \left[ \vphantom{\half}
	A_{\rho_a\rho_c}(t,\k) \, A_{\rho_b\rho_d}(t,-\k)
	+
	A_{\rho_a\rho_d}(t,\k) \, A_{\rho_b\rho_c}(t,-\k)
    \right]
\nonumber\\ &\sim&
    \chi^{1/2}_{aa'}
    \chi^{1/2}_{bb'}
    \chi^{1/2}_{cc'}
    \chi^{1/2}_{dd'} \>
    \frac{
	\delta_{a'c'} \, \delta_{b'd'} + \delta_{a'd'} \, \delta_{b'c'}
    }{
	[4 \pi (D_{a'} {+} D_{b'}) |t|]^{3/2}
    }
    \,.
\label {eq:4rhos}
\end {eqnarray}
In writing the result (\ref {eq:4rhos}),
we have chosen to use a basis for conserved charges in which
the symmetric matrix $\chi^{-1/2} D \chi^{1/2}$ is diagonal
and has eigenvalues $\{ D_a \}$.

Putting everything together yields
\begin {eqnarray}
    \volume^{-1}
    \left\langle \half \left\{ T^{ij}(t), T^{kl}(0) \right\} \right\rangle
    &=&
    \delta_{ij} \delta_{kl} \, T \, \bar\enthalpy \, v_s^2
    +
	\half \, \delta_{ij} \, \delta_{kl} \,
	\sum_{a,b}
	\frac{
	    \left[ (\chi^{1/2} \, \Xi \, \chi^{1/2})_{ab} \right]^2
	}{
	    [4\pi (D_a{+}D_b) |t|]^{3/2}
	}
\nonumber\\ && {}
    +
	\frac{2T^2 }{ 15} \,
	(
	    \half \, \delta_{ik} \delta_{jl}
	  + \half \, \delta_{il} \delta_{jk}
	  - \coeff 13 \, \delta_{ij} \delta_{kl}
	)
	\!\left[
	    \frac{1 }{ (4\pi \gamma_s |t|)^{3/2}}
	    +
	    \frac{7 }{ (8\pi \gamma_\eta |t|)^{3/2}}
	\right]\!
\nonumber\\ && {}
    +
	\frac{T^2 }{ 9} \, \delta_{ij} \, \delta_{kl}
	\left[
	    \frac{
		(1 - 3\vs^2 + \coeff 32 \vs^{-2} \xi \bar\enthalpy)^2
	    }{
		(4\pi \gamma_s |t|)^{3/2}
	    }
	+
	    \frac{
		4 \, (1 - 3\vs^2)^2
	    }{
		(8\pi \gamma_\eta |t|)^{3/2}
	    }
	\right] ,
\label {eq:stresstail}
\end {eqnarray}
up to terms vanishing faster than $|t|^{-3/2}$ as $|t| \to \infty$.

\newpage

\sloppy

\end{document}